\documentclass[twocolumn,pre,aps,showpacs]{revtex4}

\usepackage[dvips]{graphicx}
\usepackage{subfigure}
\usepackage{epsfig}

\begin{document}
\title{Correlation Effects on the Temperature Relaxation Rates in Dense Plasmas}
\author{J\'er\^ome \surname{Daligault}}\email{daligaul@lanl.gov}
\author{Guy \surname{Dimonte}}
\affiliation{Los Alamos National Laboratory, Los Alamos, New Mexico 87545}
\date{\today}

\begin{abstract}
We present a model for the rate of temperature relaxation between electrons and ions in plasmas.
The model includes self-consistently the effects of particle screening, electron degeneracy and correlations between electrons and ions.
We successfully validate the model over a wide range of plasma coupling against molecular-dynamics simulations of classical plasma of like-charged electrons and ions.
We present calculations of the relaxation rates in dense hydrogen and show that, while electron-ion correlation effects are indispensable in classical, like-charged plasmas at any density and temperature, quantum diffraction effects prevail over e-i correlation effects in dense hydrogen plasmas.
\end{abstract}

\maketitle

\section{Introduction}

Temperature relaxation rates between electrons and ions is one of many quantities that must be modeled accurately in order to predict inertial confinement fusion (ICF) \cite{Atzeni}. 
The lack of equilibrium in ICF occurs because the fusion alpha particles deposit their energy collisionally to the electrons and ions at different rates depending on the temperature regime \cite{Fraley}.
The task is challenging because ICF plasmas traverse complex physics regimes characterized by collective, quantum and correlation effects.
In practice, there have been attempts to splice models from the different regimes \cite{Brysk,LeeMore} but they are ad-hoc and not validated to the required accuracy.
Recently, direct many-body simulations \cite{Glosli,Jeon,MurilloDharmaWardana,DimonteDaligault} have been undertaken to validate the various models of relaxation rates.

Since the seminal works of Landau and Spitzer on weakly coupled plasmas \cite{LandauSpitzer}, a variety of developments has been made to calculate the temperature relaxation rates in plasmas.
The best established, parameter-free models have considered either weakly coupled, non-degenerate (ideal) plasmas (e.g. \cite{KiharaAono,BPS}) or include degeneracy effects in the limit of weak electron-ion interactions \cite{Hazak}.
In spite of recent works \cite{DharmaWardanaPerrot}, the effect of particle correlations on the electron-ion energy exchanges in non-ideal plasmas is still not definitely understood

In this paper, we present a model for the e-i temperature relaxation rates in plasmas that includes {\it self-consistently} the effects of screening, electron degeneracy and correlations between electrons and ions.
The model resolves the close and distant particle encounters in a self-consistent fashion and does not involve ad-hoc cutoffs.
We validate the model against molecular-dynamics (MD) simulations of classical plasmas over a wide range of plasma coupling.
We then apply the model to dense hydrogen and discuss the relative importance of degeneracy and correlation effects to the relaxation rate.

The paper is organized as follows.
The model is presented in section \ref{section2}.
Our derivation intentionally focusses on the temporal evolution of the {\it ionic} temperature $T_{i}$ in a two-temperature plasma.
Taking advantage that ions are classical, an equation of evolution for $T_{i}$ is obtained from momentum integration of the exact kinetic equation for the ionic phase-space distribution function.
The equation obtained (Eq.(\ref{generaldTdt}) below) expresses the change in $T_{i}$ in terms of the net work done by the electrons on the ions.
The latter can be obtained from the ionic and electronic density fluctuations in the plasma.
We thus propose a model for the density fluctuations that includes self-consistently the effects of screening, electron degeneracy and correlations between electrons and ions.
Using this model, the equation for the ionic temperature becomes a simple rate equation,
\begin{eqnarray*}
\frac{d T_{i}}{dt}&=&-\nu_{ie}\left(T_{i}-T_{e}\right)\/,
\end{eqnarray*}
where the temperature relaxation rate $\nu_{ie}$ has the Landau-Spitzer form $\nu_{ie}=\nu_{0}\ln\Lambda$ where $\nu_{0}$ is a ``universal'' rate and $\ln\Lambda$ is the generalized Coulomb logarithm that carries the many-body effects taking place in the plasma.
The model reduces to well-known approximations in the appropriate limits (e.g. Landau-Spitzer formula, Fermi-golden rule formula, Lenard-Balescu formula.)
Several technical details of the derivation of the model are presented in the appendices.

In section \ref{section3} we proceed to validate the predictions of the model by comparing them with results of MD simulations.
Because first-principles simulations of real non-equilibrium plasmas are not feasible yet, the validation upon truly ab-initio calculations is not possible.
An approximation used by several authors to cope with this consists in performing MD simulations of plasmas with semi-classical potentials that mimic quantum effects in an approximative way and prevent the unphysical collapse undergone by classical electron-ion systems \cite{Glosli,Jeon,MurilloDharmaWardana}.
However such semiclassical MD calculations are no longer fully ab-initio; besides their adequacy to simulate temperature relaxation is not unquestionably established.
The only truly ab-initio simulations of plasmas that can be performed are for like-charged systems made of positively charged electrons and ions immersed in an inert, neutralizing background.
Highly accurate MD simulations of the relaxation rate in like-charged plasmas are possible and were recently reported in \cite{DimonteDaligault}.
We therefore apply our model to a plasma of like-charged electrons and ions.
This is legimate since our formalism is independent of the nature of the electrons.
Electrons can be quantum and negatively charged as in a real plasma but can also be classical and positively charged: it is only when one evaluates the model that quantities pertaining to the system studied must be used (e.g. classical vs quantum response functions.)
As we shall see, like-charged systems are as challenging as real plasmas for the theory because, contrary to the latter, they exhibits correlation effects in all regimes of plasma coupling.
They therefore provide a good test of the validity of a theory that aims to include correlation and strong coupling effects.
Moreover, the points of differences between like-charged and real plasmas make the comparison very instructive to shed lights on correlation effects.
For classical like-charged plasmas, our model reproduces the accurate MD data and joins the weakly to strongly coupled regimes.

In section \ref{section4}, we apply the model to real dense hydrogen plasmas.
We shall see that, while e-i correlation effects are crucial in classical plasmas at any density and temperature, quantum diffraction effects prevail over e-i correlation effects in real, dense plasmas.

Finally, we present in appendix \ref{comparison} a systematic comparison of our model with the coupled-mode theory of Dharma-wardana and Perrot \cite{DharmaWardanaPerrot}, the Fermi-golden rule result \cite{Hazak,DharmaWardanaPerrot} and with the Lenard-Balescu formula \cite{Gericke}.

In the following, the Fourier transform of a space- and time-dependent function $a({\bf r},t)$ is defined as $a({\bf k},\omega)=\int_{V}{d{\bf r}\int_{-\infty}^{\infty}{dt e^{-i({\bf k}\cdot{\bf r}-\omega t)}a({\bf r},t)}}$, where $V$ is the volume.

\section{The model} \label{section2}

We consider a spatially uniform, un-magnetized, two-temperature plasma consisting of one species of ions (mass $m_{i}$, number density $n_{i}$, charge $Ze$, temperature $T_{i}$) and free electrons ($m_{e}$, $n_{e}\!=\!Zn_{i}$, $Z_{e}e$, $T_{e}$) in a volume $V$.
We assume that, at any time $t$, the electronic (e) and ionic (i) components of the plasma each may be characterized by a temperature $T_{e}(t)$ and $T_{i}(t)$, respectively.
Because of the large difference between $m_{e}$ and $m_{i}$, it is indeed reasonable to assume that the energy exchange between electrons and ions occurs on a time scale that is much larger that the equilibration times $\tau_{\alpha}$ within each subsystem $\alpha=e,i$.

The (classical) ion dynamics can be completely described by the kinetic equation for the single-particle distribution function $f_{i}({\bf r},{\bf p},t)$ in the phase space consisting of the position ${\bf r}$ and momentum ${\bf p}$ \cite{IchimaruVol1},
\begin{eqnarray}
\frac{\partial f_{i}}{\partial t}=-\frac{\partial}{\partial {\bf p}}\cdot\left\langle\delta N_{i}\,\delta{\bf F}_{i}\right\rangle\/. \label{KineticEquation}
\end{eqnarray}
Equation (\ref{KineticEquation}) is readily obtained as the ensemble average $\langle\dots\rangle$ of the evolution equation for the microscopic (Klimontovich) distribution function
\begin{eqnarray*}
N_{i}({\bf r},{\bf p};t)&=&\sum_{a=1}^{N_{i}}{\delta\left({\bf r}-{\bf r}_{a}(t)\right)\delta\left({\bf p}-{\bf p}_{a}(t)\right)}\/,
\end{eqnarray*}
where ${\bf r}_{a}(t),{\bf p}_{a}(t)$ are the position and momentum of the $a$-th ion at time $t$.
Since we are looking at time-scales larger than the equilibration times $\tau_{e,i}$ within each subsystem, $\langle\dots\rangle$ denotes an average not only over microscopic replica of the same macroscopic state but also over a time scale of order $\max(\tau_{e,i})$; the notation $\delta A$ denotes the fluctuations of a quantity $A$ around its average, i.e. $\delta A=A-\langle A\rangle$ and $\langle \delta A\rangle =0$.
In Eq.(\ref{KineticEquation}) $f_{i}=\langle N_{i}\rangle$ and $\delta {\bf F}_{i}$ is the fluctuating part of the total force acting on the ions that is induced by the density fluctuations $\delta n_{e,i}=\int{d{\bf p}\,\delta N_{e,i}}$ in the electronic and ionic systems.

The ionic temperature at time $t$ is given by,
\begin{eqnarray}
k_{B}T_{i}(t)=\frac{1}{3m_{i}n_{i}}\int{d{\bf p}\,{\bf p}^{2}f_{i}({\bf r},{\bf p},t)}\/, \label{ionictemperature}
\end{eqnarray}
Using Eqs.(\ref{KineticEquation}) and (\ref{ionictemperature}), we obtain the equation of evolution for $T_{i}$,
\begin{eqnarray}
\frac{d T_{i}}{dt}=\frac{2}{3n_{i}k_{B}}\left\langle\delta {\bf j}_{i}({\bf r},t)\cdot \delta{\bf F}_{ie}({\bf r},t)\right\rangle\/. \label{generaldTdt}
\end{eqnarray}
Here $\delta {\bf j}_{i}({\bf r},t)$ is the fluctuating part of the ionic current density $\int{d{\bf p}\,\delta N_{i}({\bf r},{\bf p},t){\bf p}/m_{i}}$.
Without magnetic fields, $\delta {\bf j}_{i}$ is longitudinal and is related to $\delta n_{i}$ through the continuity equation: $\partial \delta n_{i}({\bf r},t)/\partial t=-\nabla\cdot\delta{\bf j}_{i}({\bf r},t)$; In Fourier representation, $\delta {\bf j}_{i}=\omega \delta n_{i}({\bf k},\omega){\bf k}/k^{2}$.
The term $\delta {\bf F}_{ie}({\bf r},t)$ is the force induced by the electronic density fluctuations $\delta n_{e}({\bf r},t)$; in Fourier representation, $\delta{\bf F}_{ie}({\bf k},\omega)=i{\bf k}\/v_{ie}(k)\delta n_{e}({\bf k},\omega)$ where the e-i interaction potential is $v_{ie}$ \cite{Noteonpseudopotential}.
According to Eq.(\ref{generaldTdt}), the evolution of the ionic temperature $T_{i}$ is determined by the statistically averaged work done on the ions by the fluctuating force $\delta {\bf F}_{ie}$ exerted by the electrons

Introducing the expressions for $\delta {\bf j}_{i}({\bf k},\omega)$ and $\delta {\bf F}_{ie}({\bf k},\omega)$ in Eq.(\ref{generaldTdt}), we obtain
\begin{eqnarray}
\lefteqn{\frac{d T_{i}}{dt}=}&&\label{generaldTdt2}\\
&&\frac{2}{3n_{i}k_{B}}\frac{1}{V}\sum_{{\bf k}}{\int{d\omega\,\omega v_{ie}(k){\rm Im}\big\langle \delta n_{i}({\bf k},\omega)\delta n_{e}(-{\bf k},-\omega)\big\rangle}} \nonumber
\end{eqnarray}
where ${\rm Im}$ denotes the imaginary part.
We remark here that when the system is at equilibrium, $T_{e}=T_{i}$, the right-hand side of Eq.(\ref{generaldTdt2}) vanishes as it should.
Indeed, it follows from time-invariance that at equilibrium $\big\langle \delta n_{i}({\bf k},\omega)\delta n_{e}(-{\bf k},-\omega)\big\rangle$ is real and equal to $(2\pi)^{2}S_{ie}(k,\omega)$ where $S_{ie}(k,\omega)=1/(N_{e}N_{i})^{1/2}\int{\frac{dt}{2\pi}e^{i\omega t}\langle \delta n_{i}({\bf k},t)\delta n_{e}(-{\bf k},0)\rangle}$ is the e-i dynamic structure factor \cite{IchimaruVol1}.

According to Eq.(\ref{generaldTdt2}), an equation for $d T_{i}/dt$ can be obtained by modeling the density fluctuations in the two-temperature plasma, which we proceed to do as follows.
Following Ichimaru \cite{Ichimaru1977,IchimaruVol1}, the fluctuations $\delta N_{i}$ can generally be split into two parts, $\delta N_{i}=\delta N_{i}^{(s)}+\delta N_{i}^{(ind)}$.
$\delta N_{i}^{(s)}$ represents the spontaneous fluctuations due to the discrete nature of the particles and that are present even in the absence of interactions,
\begin{eqnarray}
N_{i}^{(s)}({\bf r},{\bf p};t)=\sum_{a=1}^{N_{i}}{\delta\left({\bf r}-{\bf r}_{a}-\frac{{\bf p}}{m_{i}}(t-t_{0})\right)\delta\left({\bf p}-{\bf p}_{a}\right)}\/,\label{Sponatneousfluctuations}
\end{eqnarray}
where ${\bf r}_{a}$ and ${\bf p}_{a}$ are the particle position and momentum at some initial time $t_{0}$.
$\delta N_{i}^{(ind)}$ describes the fluctuations that are induced by the interactions between the particles.
Upon momentum integration of Eq.(\ref{Sponatneousfluctuations}), a similar splitting holds for the electronic and ionic density fluctuations, namely
\begin{eqnarray} \label{DensityFluctuations0}
\delta n_{\alpha}({\bf r},t)=\delta n_{\alpha}^{(s)}({\bf r},t)+\delta n_{\alpha}^{(ind)}({\bf r},t)\quad\quad (\alpha=e,i)
\end{eqnarray}
where $\delta n_{\alpha}^{(s)}({\bf r},t)$ is the spontaneous fluctuations in the particle density of species $\alpha$ and $\delta n_{\alpha}^{(ind)}({\bf r},t)$ is the density fluctuations that develop due to e-e, e-i and i-i interactions.
We shall assume that the fluctuations are small quantities, i.e. $|\delta n_{\alpha}|<< n_{\alpha}$.
This is a very reasonable assumption since each subsystem $\alpha$ is supposed to be in ``internal'' equilibrium at temperature $T_{\alpha}$, and the plasma is supposed to be uniform.
Accordingly, we extend the results of equilibrium linear response theory to a two-temperature system and express the density fluctuations (\ref{DensityFluctuations0}) as \cite{Ichimaru1977,IchimaruVol1}
\begin{eqnarray}
\lefteqn{\delta n_{\alpha}(k,\omega)=\delta n_{\alpha}^{(s)}(k,\omega)}&& \label{DensityFluctuations}\\
&+&\chi_{\alpha}^{(0)}(k,\omega)\sum_{\beta=e,i}{v_{\alpha\beta}(k)\left(1-G_{\alpha\beta}(k,\omega)\right)\delta n_{\beta}(k,\omega)}\/.\nonumber
\end{eqnarray}
In Eq.(\ref{DensityFluctuations}), $v_{ab}^{eff}(k,\omega)\equiv v_{\alpha\beta}(k)\left(1-G_{\alpha\beta}(k,\omega)\right)\delta n_{\beta}(k,\omega)$ is the local effective potential seen by a particle of species $\alpha$ due to the density fluctuations $\delta n_{\beta}$, so that the product $\chi_{\alpha}^{0}({\bf k},\omega)v_{ab}^{eff}(k,\omega)$, where $\chi_{\alpha}^{0}$ is the free-particle response function of species $\alpha$, the density fluctuations resulting from the interactions of particles of species $\alpha$ with those of species $\beta$.
When $1-G_{\alpha\beta}(k,\omega)\equiv 1$, Eq.(\ref{DensityFluctuations}) corresponds to the random-phase approximation (RPA), also known as the Vlasov or mean-field approximation in plasma physics.
In this approximation, the effective field seen by an electron or an ion in the plasma is the field that would be seen by a classical, external test charge embedded in the plasma.
This approximation fails to account for the correlations that exist between the particle under scrutiny and all the other particles in the plasma.
For instance, it does not account for the fact that the effective field seen by a particle must not include the contribution from that very same particle.
The so-called local field corrections (LFC) $G_{\alpha\beta}$ in Eq.(\ref{DensityFluctuations}) account for the neglect of those correlation (and exchange) effects inherent to the RPA approximation.

The model Eq.(\ref{DensityFluctuations}) for the density fluctuations is used to calculate the right-hand-side of Eq.(\ref{generaldTdt}).
To this end, we need the correlation function for the spontaneous density fluctuations, $\langle\delta n_{\alpha}^{(s)}({\bf k},\omega)\delta n_{\beta}^{(s)}({\bf k}^{\prime},\omega^{\prime})\rangle$.
Using Eq.(\ref{Sponatneousfluctuations}) and that at $t_{0}$ the positions of two different ``spontaneous'' particles are uncorrelated, we obtain \cite{IchimaruVol1}
\begin{eqnarray} \label{spontaenousfluctuations}
\lefteqn{\langle\delta n_{\alpha}^{(s)}({\bf k},\omega)\delta n_{\beta}^{(s)}({\bf k}^{\prime},\omega^{\prime})\rangle}&&\\
&=&\delta_{\alpha\beta}(2\pi)^{2}\delta({\bf k}+{\bf k}^{\prime})\delta(\omega+\omega^{\prime})n_{\alpha}VS_{\alpha\alpha}^{(0)}(k,\omega)\nonumber
\end{eqnarray}
where $S_{\alpha\alpha}^{0}(k,\omega)=\frac{1}{N_{\alpha}}\int{\frac{dt}{2\pi}e^{i\omega\/t}\langle\delta n_{\alpha}^{(s)}(-{\bf k},0)\delta n_{\alpha}^{(s)}({\bf k},t)\rangle}$ is the dynamic structure factor of the ideal gas of species $\alpha$ at temperature $T_{\alpha}$.
The latter can be related to the free-particle response function $\chi_{\alpha}^{0}$ through the fluctuation-dissipation theorem.
\begin{eqnarray}
S_{\alpha\alpha}^{0}(k,\omega)&=&-\frac{\hbar}{\pi\/n_{\alpha}}n(\hbar\omega/k_{B}T_{\alpha}){\rm Im}\chi_{\alpha}^{0}(k,\omega) \label{fluctuationdissipation}
\end{eqnarray}
with $n(x)=1/1-e^{-x}$ for quantum particles and $n(x)=1/x$ for classical particles.
Introducing Eqs.(\ref{DensityFluctuations}), (\ref{spontaenousfluctuations}) and (\ref{fluctuationdissipation}) in Eq.(\ref{generaldTdt2}), we obtain
\begin{eqnarray} \label{dTdtgeneral2}
\lefteqn{\frac{d T_{i}}{dt}=-\frac{2\hbar}{3k_{B}\pi n_{i}V}\sum_{{\bf k}}\int{d\omega}\Big|\frac{v_{ie}(k)}{D(k,\omega)}\Big|^{2}\left[1-G_{ie}(k)\right]}&&\\
&&\times\omega\left[n(\hbar\omega/k_{B}T_{i})-n(\hbar\omega/k_{B}T_{e})\right]{\rm Im}\chi_{e}^{0}(k,\omega){\rm Im}\chi_{i}^{0}(k,\omega)\/, \nonumber
\end{eqnarray}
where
\begin{eqnarray*}
D\equiv(1-u_{ee}\chi_{e}^{0})(1-u_{ii}\chi_{i}^{0})-u_{ei}u_{ie}\chi_{e}^{0}\chi_{i}^{0}\/.
\end{eqnarray*}
and $u_{\alpha\beta}\equiv v_{\alpha\beta}(1-G_{\alpha\beta})$ \cite{NoteDetailedBalance}.
In obtaining (\ref{dTdtgeneral2}) we have assumed the static LFC approximation $G_{\alpha\beta}(k,\omega)=G_{\alpha\beta}(k,0)$, which is enough here given the additional approximation performed below; the general result is given in appendix \ref{appendix2}.

Equation (\ref{dTdtgeneral2}) can be further simplified by (a) noting that typically $m_{e}T_{i}<<m_{i}T_{e}$ and $\alpha\equiv m_{e}T_{i}/m_{i}T_{e}<<1$ can be used as a small parameter, (b) keeping the lowest order term in $\alpha$ and (c) using the f-sum rule to perform the $\omega$ integral.
The details of these operations are given in appendix \ref{appendix1}.
Equation (\ref{dTdtgeneral2}) reduces to the rate equation
\begin{eqnarray}
\frac{d T_{i}}{dt}&=&-\nu_{ie}\left(T_{i}-T_{e}\right)\/,\label{Trelaxation}
\end{eqnarray}
where the temperature relaxation rate is
\begin{eqnarray}
\lefteqn{\nu_{ie}=}&&\label{model}\\
&&-\frac{1}{3\pi^{2}\/m_{i}}\int_{0}^{\infty}{\!\!\!\!dk k^{4}\left|\frac{v_{ei}(k)}{\epsilon_{e}(k,0)}\right|^{2}\left(1\!-\!G_{ie}(k)\right)\frac{\partial {\rm Im}\chi_{e}^{0}(k,\omega)}{\partial\omega}\Big|_{\omega=0}}\/. \nonumber
\end{eqnarray}
Here $\epsilon_{e}(k,0)=1-v_{ee}(k)(1-G_{ee}(k))\chi_{e}^{0}(k,0)$ is the static electronic dielectric function, $G_{\alpha\beta}(k)=G_{\alpha\beta}(k,0)$ are the static LFC's.

Our model (\ref{model}) is conveniently rewritten in the Landau-Spitzer form 
\begin{eqnarray*}
\nu_{ie}=\nu_{0}\ln\Lambda\/,
\end{eqnarray*}
i.e as the product of a ``universal'' rate
\begin{eqnarray*}
\nu_{0}=\frac{8n_{e}Z{^2}e^{4}\sqrt{2\pi m_{e}m_{i}}}{3(m_{i}k_{B}T_{e})^{3/2}}
\end{eqnarray*}
and of a (dimensionless) generalized Coulomb logarithm,
\begin{eqnarray}
\!\!\!\ln\Lambda&=&-\sqrt{\frac{2}{\pi m_{e}}}\frac{(k_{B}T_{e})^{3/2}}{n_{e}}\frac{1}{(4\pi Ze^{2})^{2}}\label{modellnL}\\
&&\times\int_{0}^{\infty}{\!\!\!\!dk k^{4}\left|\frac{v_{ei}(k)}{\epsilon_{e}(k,0)}\right|^{2}\left(1\!-\!G_{ie}(k)\right)\frac{\partial {\rm Im}\chi_{e}^{0}(k,\omega)}{\partial\omega}\Big|_{\omega=0}}\/. \nonumber
\end{eqnarray}

Our approach, which consists in modeling the density fluctuations in a two-temperature plasma to obtain $dT_{i}/dt$, does not rely on the concept of isolated binary collisions and allows us to treat the plasma as a single entity and to include self-consistently the electron-ion interactions.
For instance the collective behavior typical of a plasma and in particular the screening of the e-i interaction due to both electrons and ions are present through the dielectric function $D(k,\omega)$ in Eq.(\ref{dTdtgeneral2}).
The short-range correlations, which especially affect the contribution of close encounters, are self-consistently added through the local-field corrections (see e.g. the factor $1-G_{ie}$ in the numerator of Eqs.(\ref{dTdtgeneral2}) and \ref{modellnL}.)
The effect of particle statistics, e.g. electron degeneracy, is included through the response functions.
All these effects are not {\it ad hoc} constructs but are self-consistently derived.
Their contributions are analysed in the next sections.

A detailed comparison of our model with others models such as the Fermi-golden rule formula \cite{Hazak}, the Lenard-Balescu formula \cite{Gericke,DaligaultMozyrsky} and the coupled-mode theory of Dharma-wardana and Perrot \cite{DharmaWardanaPerrot} is presented in appendix \ref{comparison}.
Here we just make the following remarks.
When only e-i correlations are neglected, $G_{ie}=0$, and $G_{ee}$ is approximated by its value in the jellium model $G_{ee}^{jel}$, our model reduces to the so-called Fermi-golden rule (FGR) formula obtained within the framework of linear response theory assuming weak e-i interactions \cite{Hazak}.
Our model can thus be regarded as an extension of the FGR formula where the plasma is treated as a whole and e-i interactions are treated self-consisitently.
When the LFC's are completely neglected, $G_{\alpha\beta}=0$, our model reduces to the result obtained using the Lenard-Balescu kinetic equations.

Finallty, it is worthwhile to remark that the Coulomb logarithm (\ref{modellnL}) differs from the Coulomb logarithm entering the Ziman formula for the electronic conductivity and its extensions to strongly coupled plasmas \cite{IchimaruVol1,IchimaruTanaka1985,Boercker} in that the latters involve, in addition to the terms in the integrand of Eq.(\ref{modellnL}), the static ion-ion structure factor $S_{ii}(k)$.
Thus, contrary to the e-i momentum exchanges that govern the electronic conductivity, e-i energy exchanges are globally insensitive to the details of the ion fluctuation spectrum \cite{Hazak} (see discussion in appendix \ref{appendix1}.)
This is reminiscent of Bethe's result on stopping power \cite{Bethe1930,IchimaruVol1} that the energy loss of a fast charged particle in a plasma is fixed by the total number of scatterers.

\section{Validation of the model} \label{section3}

\begin{figure}[t]
\includegraphics[scale=0.50]{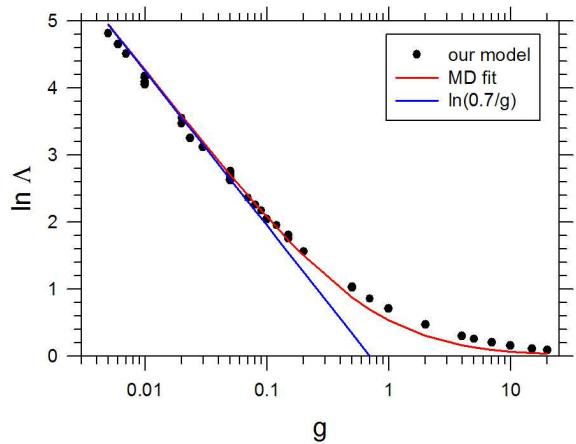}
\caption{\label{figure1}
(color) Coulomb logarithm vs plasma parameter for like-charged systems. The red line is a fit to the MD data \cite{DimonteDaligault}.}
\end{figure}
We validate Eq.(\ref{model}) using ab-initio MD simulations of classical e-i systems with a pure Coulomb potential.
We avoid Coulomb collapse (recombination) by using positively (like-) charged electrons and ions immersed in an inert, neutralizing background.
As explained in the introduction, this is done because fully quantum mechanical simulations are not yet feasible.
Again, this is legimate since our theory can {\it also} be applied to classical like-charged systems.
In the classical limit (see appendix \ref{appendixChi}), we have
\begin{eqnarray*}
\frac{\partial}{\partial \omega}{\rm Im}\chi_{e}^{0}(k,0)&=&-\frac{n_{e}}{(k_{B}T_{e})^{3/2}}\sqrt{\frac{\pi m_{e}}{2}}\frac{1}{k} \label{classicalChi}\\
\chi_{e}^{0}(k,0)&=&-n_{e}/k_{B}T_{e}\/,
\end{eqnarray*}
and Eq.(\ref{model}) becomes,
\begin{eqnarray} \label{classicalmodel}
\ln \Lambda=\int_{0}^{\infty}{\frac{dk}{k}\frac{1-G_{ie}(k)}{|\epsilon_{e}(k,0)|^{2}}}\/,
\end{eqnarray}
where $\epsilon_{e}(k,0)=1+(k_{De}^{2}/k^{2})(1-G_{ee}(k))$ and $k_{De}=(4\pi n_{e}e^{2}/k_{B}T_{e})^{1/2}$ is the inverse electronic Debye length.

The Coulomb logarithm $\ln\Lambda$ for temperature relaxation in a like-charged, classical plasma ($Z_{e}=Z=1$) is shown in Fig.\ref{figure1} from various calculations as a function of the plasma coupling parameter $g=r_{L}/\lambda_{De}$ where $r_{L}=e^{2}/k_{B}T_{e}$ is the distance of closest approach (Landau length) and $\lambda_{De}=1/k_{De}$ is the electron Debye length.
The red line is a fit to results from accurate, large-scale MD simulations \cite{DimonteDaligault}.
For $g<<1$, the MD simulations confirm the theories \cite{KiharaAono,BPS}, which regularize the divergent collision integrals at small and large momentum transfer $k$, and yield $\ln(0.765\,\lambda_{De}/r_{L})$ (blue line).
However, these theories breakdown at $g>0.1$, since they do not satisfactorily describe correlation effects and $\ln\Lambda<0$.
Our model (\ref{classicalmodel}), whose results are shown by the black dots in Fig.(\ref{figure1}), not only recovers the weak coupling limit $\ln(0.765 \lambda_{De}/r_{L})$ but also is in very good agreement with the MD calculations over the whole range of coupling.
\begin{figure}[t]
\includegraphics[scale=0.50]{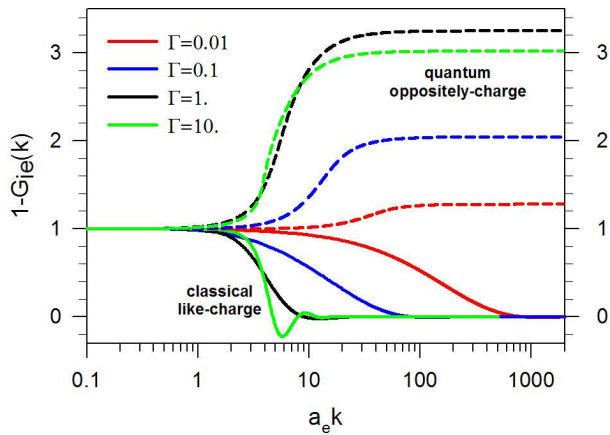}\\
\caption{\label{figure4}
(color) $1-Gie(k)$ (on a semi-log scale) for classical, like-charged hydrogen (lower, full curves) and for real hydrogen with quantum, negative electrons (upper, dashed curves) at $n_{e}=1.6\,10^{24}$ $\rm cm^{-3}$ and for $\Gamma=0.01,0.1,1.$ and $10.$.}
\end{figure}

\begin{figure}[t]
\includegraphics[scale=0.51]{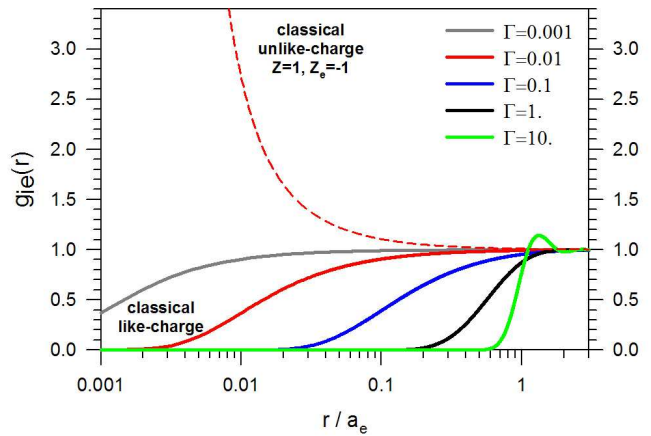}\\
\caption{\label{figure2}
(color) Electron-ion pair distribution function (on a semi-log scale) for like-charged hydrogen (full lines) at $\Gamma=0.001,\/0.01,\/0.1,\/1$ and $10$. The dashed line shows $g_{ie}(r)$ for a purely classical plasma with negatively-charged electrons at $\Gamma=0.01$. Here $\Gamma=e^{2}/a_{e}k_{B}T_{e}=(g/\sqrt{3})^{2/3}$ is used to characterize the plasma coupling., where $a_{e}=(3/4\pi\/n_{e})^{1/3}$ is the mean interparticle distance.
}
\end{figure}

Correlations effects are accounted for by the LFC term $1-G_{ie}$.
When correlations are neglected ($G_{\alpha\beta}=0$), equation (\ref{classicalmodel}) diverges logarithmically at large momentum $k$ for all plasma coupling,
\begin{eqnarray*}
\ln \Lambda=\int_{0}^{\infty}{dk\,k^{3}/(k^{2}+k_{De}^{2})^{2}}=\infty\/\quad\forall g\/.
\end{eqnarray*}
The integral diverges due to close encounters (large $k$) because the neglect of correlations assumes that the pair-distribution functions $g_{\alpha\beta}(r)\sim 1$ everywhere.
However, as illustrated in Fig.(\ref{figure2}), repulsion at small inter-particle distances $r$ forces $g_{\alpha\beta}(r)$ to vanish continuously at $r=0$.
In our model, we account for the ``hole'' in $g_{ab}(r)$ by the term $1-G_{ie}(k)$ since the two quantities are related by the Ornstein-Zernicke (OZ) relation between the pair-distribution functions and the direct correlation functions $c_{\alpha\beta}(k)=-v_{\alpha\beta}(k)(1-G_{\alpha\beta}(k))/k_{B}T$ \cite{IchimaruVol1}.
In particular, the OZ relation implies \cite{NoteOZ}
\begin{eqnarray}
1-G_{ie}(k)&=&\frac{Z_{e}}{4\pi\/e^{2}\sqrt{Z}}\frac{D(k,0)}{\chi_{e}^{0}(k,0)}k^{2}S_{ie}(k) \label{Gie}
\end{eqnarray}
where $S_{ie}(k)=\sqrt{n_{e}n_{i}}\int{d{\bf k}(g_{ie}(r)-1)e^{-i{\bf k}\cdot{\bf r}}}$ is the i-e structure factor.
We calculate $G_{\alpha\beta}(k)$ and $g_{\alpha\beta}(r)$ self-consistently by using the hypernetted chain (HNC) closure $g_{\alpha\beta}(r)=\exp\left(-v_{\alpha\beta}/k_{B}T+g_{\alpha\beta}-1-c_{\alpha\beta}\right)$ in the OZ relations.
The HNC closure is known to accurately describe correlations in classical Coulomb systems \cite{IchimaruVol1,noteHNC}.
Results for $g_{ie}(r)$ and $1-G_{ie}(k)$ are shown in Figs.(\ref{figure2}) and (\ref{figure4}) for various values of coupling.
As a consequence of short-range correlations and for all coupling parameters,
\begin{eqnarray*}
g_{ie}(r=0)=0
\end{eqnarray*}
and Eq.(\ref{Gie}) implies (see appendix \ref{appendixlimGie})
\begin{eqnarray}
\lim_{k\to 0}{1-G_{ie}(k)}=0\/. \label{limGieclassical}
\end{eqnarray}
As a consequence Eq.(\ref{classicalmodel}) converges for all plasma coupling,
\begin{eqnarray*}
\ln \Lambda=\int_{0}^{\infty}{\frac{dk}{k}\frac{1-G_{ie}(k)}{|\epsilon_{e}(k,0)|^{2}}}\,<\,\infty\/\quad\forall g.
\end{eqnarray*}
Thus, while Debye screening cuts off the integral (\ref{classicalmodel}) at small $k$ (distant encounters), short-range e-i correlation effects embodied in $1-G_{ie}(k)$ are crucial to provide the large momentum cutoff.
The results obtained for $\ln\Lambda$ (black dots in Fig.(\ref{figure1})) are in very good agreement with the MD calculations over the whole range of coupling.
At small $g$, Eq.(\ref{classicalmodel}) recovers the weak-coupling limit $\ln(0.765 \lambda_{De}/r_{L})$.
At higher coupling, it joins the weakly to strongly coupled regimes.
We also find (not shown here) that $\ln\Lambda$ is insensitive to the ion charge $Z$ at constant $g$, consistent with MD simulations \cite{DimonteDaligault}.

\begin{figure}[t]
\includegraphics[scale=0.45]{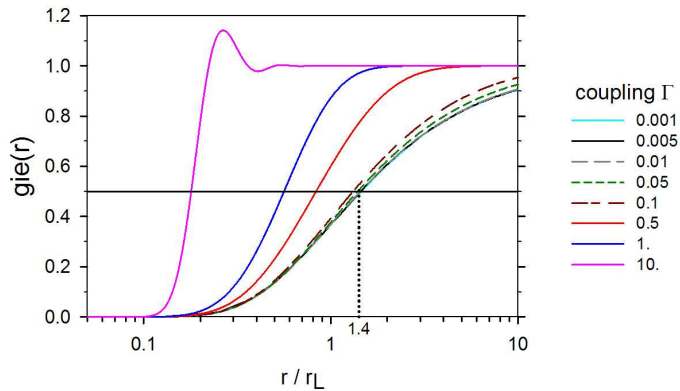}
\caption{\label{figure3}
(color) Electron-ion pair distribution functions (on a semi-log scale) for like-charged hydrogen as a function of $r/r_{L}$ where $r_{L}$ is the Landau length for and for coupling $0.001\leq\Gamma=e^{2}/a_{e}k_{B}T_{e}\leq 10.$. It is remarkable that, for all $\Gamma\leq 0.1$, $g_{ie}(r)=0.5$ when $r\approx 1.4 r_{L}$.}
\end{figure}
It is remarkable that Eq.(\ref{classicalmodel}) recovers the result of Kihara and Aono (KH) \cite{KiharaAono} and rederived recently with modern regularization techniques by Brown et al. (BPS) \cite{BPS} (see also \cite{LandauLifshitz}.)
Both KH and BPS use sophisticated regularization techniques to eliminate the divergences encountered when dealing with Coulomb collisions in a plasma with the traditional particle and wave pictures. 
In the particle picture, particles undergo binary Coulomb collisions and the dynamics of each charged species is governed by a Boltzmann equation with Rutherford cross sections.
The collision integrals, expressed as integrals over the impact parameter, diverge logarithmically at large impact parameter because collective interactions between charged particles are not included.
In the wave picture, those collective (screening) effects are included through the plasma dielectric function.
In the traditional approach of Landau, large momentum transfer are neglected and particles momenta diffuse in the fluctuations of the total electric field of the plasma.
As a consequence, close collisions are not treated correctly and the collision integrals diverge at large momentum transfer.
At first sight, the agreement between our model for like-charged plasmas and \cite{KiharaAono,BPS} might seem curious since these authors consider real plasmas with negatively charged electrons in the so-called classical regime where $r_{L}>\lambda_{th}$, where $\lambda_{th}=\hbar/\sqrt{m_{e}k_{B}T_{e}}$ is the thermal de Broglie wavelength.
This is because the fundamental quantities underlying KH and BPS theories, namely the Rutherford cross section and the dielectric function \cite{KiharaAono,BPS}, scale like the square of the charges and their results are therefore insensitive to the sign of the electron charge.

Following Landau and Spitzer's original works \cite{LandauSpitzer}, we note that the result $\ln(0.765 \lambda_{De}/r_{L})$ can be interpreted in terms of maximum and minimum impact parameter cutoffs by writing $\ln\Lambda=\ln(b_{max}/b_{min})$ with $b_{max}=\lambda_{De}$ and $b_{min}=r_{L}/0.765\approx 1.4 r_{L}$, respectively.
$b_{max}$ is the widespread used maximum cutoff that arises from electronic screening while $b_{min}$ corresponds to the distance below which $g_{ie}(b_{min})<0.5$.
Figure (\ref{figure3}) indeed shows that at small coupling $g_{ie}(r)$ is always equal to $0.5$ when $r\approx 1.4 r_{L}$.
Our approach therefore allows us to understand the statistical origin of $b_{min}$: close collisions below that distance are statistically rare and do not contribute to the Coulomb logarithm.

Finally, our model (\ref{classicalmodel}) diverges when applied to fully classical and negatively-charged electron; indeed, as illustrated in Fig.\ref{figure2}, since $g_{ie}(r)\sim\exp(-v_{ei}(r)/k_{B}T_{e})=\exp(-Z_{e}Z/r\/k_{B}T_{e})\to \infty$ at $r=0$, $\lim_{k\to\infty} 1-G_{ie}(k)=-\infty$ at all coupling when $Z_{e}=-1.$, indicating the infinite attraction between classical ions and classical, point-like electrons.
On the contrary, KH and BPS converge in the classical limit because the Rutherford cross section is independent of the particle distributions, and in particular of $g_{ie}(r)$.
In our model, negatively-charged electrons must be treated quantum-mechanically, as shown in the next section.

\section{Application to Dense Hydroden plasmas}  \label{section4}

\begin{figure}[t]
\includegraphics[scale=0.51]{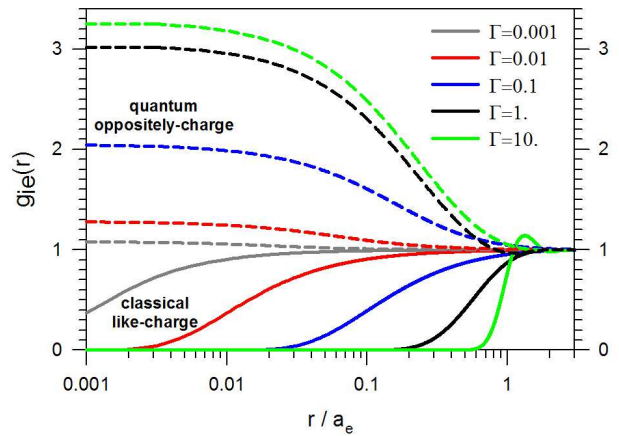}\\
\caption{\label{figure5}
(color) Electron-ion pair distribution functions (on a semi-log scale) for like-charged hydrogen (lower, full curves) and for real hydrogen plasmas with quantum, negative electrons (upper, dashed curves) at $n_{e}=1.6\,10^{24}$ $cm^{-3}$ and for $\Gamma=0.01,0.1,1$ and $10$.}
\end{figure}
Having validated our model for classical like-charged plasmas, we now consider fully ionized (hydrogen) plasmas with negative  ($Z_{e}=-1$), quantum mechanical electrons.
We define the usual coupling parameter $\Gamma=e^{2}/a_{e}k_{B}T_{e}$ where $a_{e}=(3/4\pi\/n_{e})^{1/3}$ is the mean interparticle distance, the degeneracy parameter $\Theta=k_{B}T_{e}/E_{F}$ where $E_{F}=\hbar^{2}k_{F}^{2}/2m_{e}$ and $k_{F}=(3\pi n_{e})^{1/3}$ are the electronic Fermi energy and momentum, and $r_{s}=a_{e}/a_{B}$ where $a_{B}$ is the Bohr radius.
Using the quantum expression for $\frac{\partial}{\partial \omega}{\rm Im}\chi_{e}^{0}(k,0)$ (see appendix \ref{appendixChi}),
\begin{eqnarray}
\frac{\partial}{\partial \omega}{\rm Im}\chi_{e}^{0}(k,0)&=&-\frac{n_{e}}{(k_{B}T_{e})^{3/2}}\sqrt{\frac{\pi m_{e}}{2}}\frac{1}{k}f(k/2) \label{quantumImChip}
\end{eqnarray}
where $f(k)=\frac{3\sqrt{\pi}}{4}\,\Theta^{3/2}[1+e^{(k^{2}/k_{F}^{2}-\mu/E_{F})/\Theta}]^{-1}$, simply adds a Fermi distribution factor $f(k/2)$ to the classical expression used in the previous section.
Our model (\ref{model}) becomes,
\begin{eqnarray}\label{quantummodel}
\ln \Lambda=\int_{0}^{\infty}{\frac{dk}{k}\frac{1-G_{ie}(k)}{|\epsilon_{e}(k,0)|^{2}}\/f(k/2)}\/.
\end{eqnarray}

\begin{figure}[t]
\includegraphics[scale=0.50]{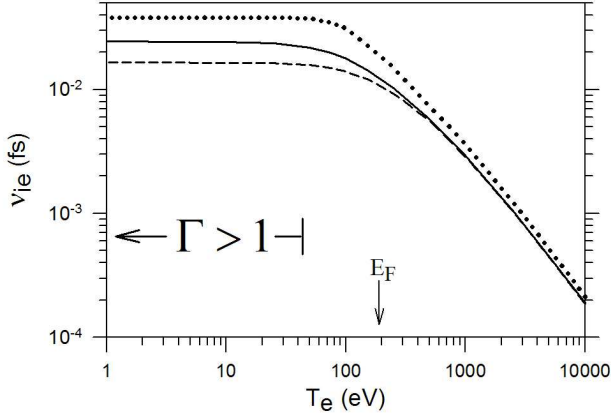}\\
\caption{\label{figure6}
Relaxation rate $\nu_{ie}$ for dense hydrogen at $n_{e}=1.3e25$ $\rm g.cm^{-3}$ obtained using (\ref{quantummodel}) (full line), the FGR with $G_{ee}=G_{ii}=0$ (dashed lined), and the widely used Brysk formula \cite{Brysk} (dotted line).}
\end{figure}
Equation (\ref{quantummodel}) differs from the classical limit (\ref{classicalmodel}) in two major ways.
{\it First}, it converges even when correlations are neglected, i.e. $G_{\alpha\beta}=0$, because $f(k/2)$ vanishes exponentially at large $k$ and cuts off the integral for $k$ of the order of $2k_{F}\sqrt{1+\Theta}$,
\begin{eqnarray}
\int_{0}^{\infty}{\frac{dk}{k}\frac{1}{|\epsilon_{e}(k,0)|^{2}}\/f(k/2)}<\infty \quad \forall\Gamma\,,\,\forall\Theta \label{quantummodelGiezero}
\end{eqnarray}
For instance, in the non-degenerate limit $\Theta >>1$, Eq.(\ref{quantummodelGiezero}) gives
\begin{eqnarray}
\ln\Lambda&=&\int_{0}^{\infty}{dk\,\frac{k^{3}}{(k^{2}+k_{De}^{2})^{2}}e^{-k^{2}/4k_{F}^{2}\Theta}} \quad (\Theta>>1) \nonumber\\
&\approx& \ln(0.742\sqrt{r_{s}}/\Gamma) \label{lnLqLB}
\end{eqnarray}
This result corresponds to that obtained using the quantum Lenard-Balescu kinetic equation \cite{DaligaultMozyrsky}.
It was more recently derived by BPS \cite{BPS} and can also be found in \cite{LandauLifshitz}.
Following Landau and Spitzer, Eq.(\ref{lnLqLB}) can be rewritten as $\ln\Lambda=\ln(b_{max}/b_{min})$ in terms of the maximum and minimum impact parameters $b_{max}=\lambda_{De}$ and $b_{min}\approx 0.778\,\lambda_{th}$, where $\lambda_{th}=\hbar/\sqrt{m_{e}k_{B}T_{e}}$ is the thermal de Broglie wavelength; this results is to be compared to $b_{min}=1.4\,r_{L}$ obtained in the previous section for classical like-charged systems.

In the degenerate limit $\Theta<<1$, $f(k/2)\approx\theta(2k_{F}-k)$ and the range of integration is limited to $2k_{F}$ (only those electrons near the Fermi surface take part in energy exchanges),
\begin{eqnarray}
\ln \Lambda&=&\int_{0}^{2k_{F}}{\frac{dk}{k}\frac{1}{|\epsilon_{e}(k)|^{2}}}\nonumber\\
&\approx&\frac{1}{2}\left[\ln\left(1+\frac{4k_{F}^{2}}{k_{TF}^{2}}\right)-\frac{4k_{F}^{2}}{4k_{F}^{2}+k_{TF}^{2}}\right]\/, \label{extendedBrysk}
\end{eqnarray}
For illustration, we used in the last equation the large wavelength appoximation $\epsilon_{e}(k)\approx 1+k_{TF}^{2}/k^{2}$ where $k_{TF}$ is the finite temperature Thomas-Fermi screening length \cite{NotekTF}.
By comparing Eq.(\ref{extendedBrysk}) with the Brysk formula \cite{Brysk,LeeMore}, we remark that Brysk et al. used only the logarithmic part.
At high density, the second term makes a negative correction to the logarithmic part of typically $20-30\%$.

As shown in appendix \ref{appendixChi}, the presence of the converging factor $f(k/2)$ in Eq.(\ref{quantumImChip}) and its absence in Eq.(\ref{classicalChi}) are due to the difference in the energy excitations $\omega={\bf k}\cdot{\bf p}/m_{e}+\hbar\/k^{2}/2m_{e}$ and $\omega={\bf k}\cdot{\bf p}/m_{e}$, respectively.
In particular, in the quantum electron gas and for the small energy transfers $\omega\sim 0$ of interest here, large momentum transfers ($k^{2}>> m_{e}k_{B}T_{e}/\hbar^{2}$) characteristic of close encounters can only involve electrons in the tail of the Fermi distribution.
Since the latter vanishes exponentially at large momentum, the cumulating effects of recoil energy and Fermi statistics, hereafter referred to as quantum diffraction effect, naturally cuts off Eq.(\ref{quantummodel}) at large $k$ irrespective of the strength of the e-i correlations.

{\it Second}, as illustrated in Figs.(\ref{figure4}) and (\ref{figure5}) for dense hydrogen, real plasmas exhibit correlation effects that differ from like-charged plasmas.
Here $g_{ie}(r)$ varies from 
\begin{eqnarray*}
g_{ie}(r=0)>1\/,
\end{eqnarray*}
at the origin to $g_{ie}(r)=1$ at large distance $r$; moreover, $g_{ie}(r)\to 1$ everywhere as the temperature increases.
The quantity $1-G_{ie}(k)$ still obeys the OZ relation Eq(\ref{Gie}) \cite{Chihara}, and varies from $1$ at $k=0$ to (see appendix \ref{appendixlimGie})
\begin{eqnarray}
\lim_{k\to 0}{1-G_{ie}(k)}\propto g_{ie}(r=0)> 0\/, \label{limGiequantum}
\end{eqnarray}
at large $k$, as illustrated in Fig.\ref{figure4}.
Thus, in contrast with the classical limit, $1-G_{ie}(k)$ does not cutoff the integral (\ref{quantummodel}) at large $k$.
The cutoff is instead provided by the Fermi distribution $f(k/2)$.
Moreover, the calculations of \cite{ChiharaGei,Tanaka,NagaoBonevAshcroft} suggest that $G_{ie}(k)\leq 0$ and $G_{ee}(k)\geq 0$, and therefore Eq.(\ref{quantummodel}) suggests that correlations tend to increase the Coulomb logarithm from its value obtained assuming $G_{\alpha\beta}=0$, in agreement with the conclusions of \cite{DaligaultMozyrsky}.

We estimated the LFC's needed in Eq.(\ref{quantummodel}) following the approach prescribed in \cite{NagaoBonevAshcroft} for {\it dense} hydrogen plasmas.
For all the densities studied ($10^{23}\leq n_{e}\leq 10^{27}$ $\rm cm^{-3}$), a similar behavior illustrated in Fig.(\ref{figure6}) is observed.
At small temperature, $\nu_{ie}$ stays nearly constant up to $k_{B}T_{e}\approx 0.5E_{F}$ at a value slightly higher than when correlations are neglected \cite{DaligaultMozyrsky}; indeed, $\ln\Lambda$ scales like $T_{e}^{3/2}$ at small $T_{e}$, which cancels with the $T_{e}^{-3/2}$ in $\nu_{0}$.
Beyond  $k_{B}T_{e}> 0.5E_{F}$, the rate decreases and at high temperature it follows the quantum Lenard-Balescu result $\nu_{0}\ln\Lambda$ with $\ln\Lambda$ given by Eq.(\ref{lnLqLB}), indicating that, in contrast with the classical system studied above, correlations do not play any important role.
Here, quantum diffraction effects play a bigger role than e-i correlation effects in determining $\nu_{ie}$, and the FGR formula decently estimates the relaxation rates.
Note that at these high densities, electron degeneracy is always important when the plasma coupling is large than unity, i.e. $\Theta< 1$ when $\Gamma>1$.
We expect that correlations will play a bigger role whenever $g_{ie}(r=0)$ (and in turn $|1-G_{ie}(\infty)|\propto g_{ie}(r=0)$) significantly increases while the large momentum cutoff $2k_{F}\sqrt{1+\Theta}$ imposed by $f(k/2)$ also increases; this certainly occurs at densities and temperatures low enough for bound states to emerge and below which our model breaks down \cite{NagaoBonevAshcroft}.

\section{Conclusion}

In summary, we describe a model for the rate of temperature relaxation between electrons and ions that treats the effects of electron statistics and particle screening and correlations in a self-consistent fashion.
Such a treatment is necessary in order to calculate the Coulomb logarithm without ad-hoc cutoffs and with improved accuracy for the various physics regimes encountered in inertial confinement fusion and stellar interiors.
The key result in our model for the relaxation rate removes the uncertainty in the Coulomb logarithm because it resolves the close and distant particle encounters in a self-consistent fashion.
The distant encounters are limited by the plasma dielectric response for Boltzmann or Fermi statistics as needed.
The close encounters are limited by quantum diffraction effects and short-range particle correlations.
By treating these effects together and self-consistently, the Coulomb logarithm that we obtain for a low temperature, oppositely-charged plasma is $50\%$ smaller than the often used Brysk \cite{Brysk} result and $50\%$ larger than the Fermi Golden rule.

In order to validate our model with ab-initio MD simulations, we also applied our fomalism to a plasma of like-charged electrons and ions.
This is motivated by the fact that experiments with real plasmas have not been able to provide data of sufficient accuracy to resolve any of these issues.
The available MD simulations can be much more accurate but they are fundamentally classical.
Moreover, when the interparticle potential is modified at short distance to include to some degree quantum effects, the simulations are no longer ab-initio.
We thus abandoned the semi-classical approximation and used the real Coulomb force in our MD simulations, but we had to make the electron and ion charges alike in order to avoid Coulomb collapse.
Our model is applicable to such a system and our results are in excellent agreement with MD simulations.
This is a valid test of our formalism because the extension to a real and quantum plasma is straightforward.

In the future, we plan to perform MD simulations with modfied potentials in order to test the semiclassical approximation for both like and opposite charged plasmas \cite{DimonteDaligaultPhysPlasma}.
In addition, we will compare our model results with previous models in ICF experiments.

\begin{acknowledgments}
This work was performed for the U.S. Department of Energy by Los Alamos National Laboratory under contract DE-AC52-06NA25396.
\end{acknowledgments}

\appendix

\section{General expression} \label{appendix2}

Eq.(\ref{dTdtgeneral2}) is obtained assuming the static LFC approximation $G_{ab}(k,\omega)\equiv G_{ab}(k,0)$.
In general, we obtain the more complicated expression,
\begin{eqnarray}
\lefteqn{\frac{d T_{i}}{dt}=\frac{2}{3k_{B} n_{i}V^{2}}\sum_{{\bf k}}{\int{d\omega \frac{v_{ie}(k)}{|D(k,\omega)|^{2}}\omega}}}\label{dTdtgeneral}\\
&&\times\left[{\rm Im}A_{ei}(k,\omega)S_{ii}^{0}(k,\omega)-{\rm Im}A_{ie}(k,\omega)S_{ee}^{0}(k,\omega)\right]\nonumber
\end{eqnarray}
with $A_{\alpha\beta}=u_{\alpha\beta}\chi_{\alpha}^{0}(1-u_{\alpha\alpha}^{*}\chi_{\alpha}^{0,*})$, $D=(1-u_{ee}\chi_{e}^{0})(1-u_{ii}\chi_{i}^{0})-u_{ei}u_{ie}\chi_{e}^{0}\chi_{i}^{0}$ and $u_{\alpha\beta}\equiv v_{\alpha\beta}(1-G_{\alpha\beta})$ (the star denotes the complex conjugate.)
Eq.(\ref{dTdtgeneral}) reduces to Eq.(\ref{dTdtgeneral2}) when $G_{ab}(k,\omega)\equiv G_{ab}(k,0)$.
In that case, the LFC's are real and ${\rm Im} A_{\alpha\beta}=u_{\alpha\beta}{\rm Im}\chi_{\alpha}^{0}$.
To first order in the small parameter $m_{e}T_{i}/<m_{i}T_{i}$, both Eq.(\ref{dTdtgeneral2}) and Eq.(\ref{dTdtgeneral}) lead to the expression (\ref{model}) for $\nu_{ie}$.

Note that the right-hand side of Eq.(\ref{dTdtgeneral}) is equivalent to Eq.(\ref{generaldTdt2}) and therefore vanishes when $T_{e}=T_{i}$.

\section{Important properties of response functions} \label{appendixChi}

For completeness, we list in this appendix a number of basic properties satisfied by quantum and classical response functions.
Most of these properties can be found (sometimes with different notations) for instance in \cite{IchimaruVol1} and \cite{Kremp}.

\subsection{Classical and quantum free-particle response functions}

We consider an homogeneous system consisting of a single species of non-interacting particles of mass $m$ and characterized by the particle density $n$, the (inverse) temperature $\beta=1/k_{B}T$ and the chemical potential $\mu$.

If the system is treated quantum-mechanically, the density-density response function $\chi^{(0)}(k,\omega)$ is given by
\begin{eqnarray} \label{fullequation}
\chi^{(0)}(k,\omega)=-\int{\frac{d{\bf p}}{(2\pi)^{3}}\frac{F({\bf p}+\hbar{\bf k})-F({\bf p})}{\hbar\omega-\Delta E({\bf p},{\bf k})+i0+}}
\end{eqnarray}
where $F$ is the Fermi distribution
\begin{eqnarray}
F(p)=\frac{1}{1+\exp\beta(\epsilon({\bf p})-\mu)}\/,
\end{eqnarray}
and $\Delta E({\bf p},{\bf k})=\epsilon({\bf p}+\hbar{\bf k})-\epsilon({\bf p})$ where $\epsilon({\bf p})={\bf p}^{2}/2m$ is the energy of a particle of momentum ${\bf p}$.
Equation (\ref{fullequation}) implies
\begin{eqnarray}
{\rm Im}\chi^{(0)}(k,\omega)&=&\pi\int{\frac{d{\bf p}}{(2\pi)^{3}}\left[F({\bf p}+\hbar{\bf k})-F({\bf p})\right]}\nonumber\\
&&\times\delta\left(\hbar\omega-\Delta E({\bf p},{\bf k})\right)\label{quantumImChi}\\
&=&\frac{m^{2}}{2\pi\hbar^{4}\beta}\frac{1}{k}\ln\frac{1+e^{\beta\left(-\nu_{+}^{2}\epsilon_{F}+\mu\right)}}{1+e^{\beta\left(-\nu_{-}^{2}\epsilon_{F}+\mu\right)}}\/, \label{quantumImChi2}
\end{eqnarray}
with $\nu_{\pm}=\frac{\omega}{qv_{F}}\pm\frac{q}{2k_{F}}$.
By differentiating (\ref{quantumImChi2}) with respect to $\omega$, we find
\begin{eqnarray}
\frac{\partial}{\partial\omega}{\rm Im}\chi^{(0)}(k,\omega=0)&=&-n\beta\sqrt{\frac{\pi m\beta}{2}}\frac{1}{k}f(k/2)\/,
\end{eqnarray}
with
\begin{eqnarray}
f(k)\equiv\frac{3\sqrt{\pi}}{4}\Theta^{3/2}F(\hbar k)\/. \label{fofk}
\end{eqnarray}
In the classical limit ($\hbar\to 0$), Eq.(\ref{quantumImChi}) becomes
\begin{eqnarray}
{\rm Im}\chi^{(0)}(k,\omega)&=&n\pi\hbar\int{\frac{d{\bf p}}{(2\pi)^{3}}{\bf k}\cdot\nabla F_{cl}(p)}\nonumber\\
&&\times\delta\left(\hbar\omega-\Delta E({\bf p},{\bf k})\right) \label{classicalImChi} \\
&=&-n\beta\sqrt{\pi}Y\/e^{-Y^{2}}\/,
\end{eqnarray}
where $Y=\sqrt{m\beta/2}\omega/k$, $\Delta E({\bf p},{\bf k})=\hbar{\bf k}\cdot{\bf p}/m$ and $F_{cl}$ is the Maxwellian distribution
\begin{eqnarray}
F_{cl}(p)=\left(\frac{m\beta}{2\pi}\right)^{3/2}\exp\left(-\beta{\bf p}^{2}/2m\right)\/.
\end{eqnarray}
The frequency derivative at $\omega=0$ is
\begin{eqnarray}
\frac{\partial}{\partial\omega}{\rm Im}\chi^{(0)}(k,\omega=0)&=&-n\beta\sqrt{\frac{\pi m\beta}{2}}\frac{1}{k}\/.
\end{eqnarray}
The only change between the quantum and classical expressions for $\frac{\partial}{\partial\omega}{\rm Im}\chi^{(0)}(k,\omega=0)$ is in the factor $f(k/2)$.
The difference stems from the difference in the quantum and classical recoil energies $\Delta E({\bf p},{\bf k})$ appearing in the delta functions of Eqs.(\ref{quantumImChi}) and (\ref{classicalImChi}), viz.
\begin{eqnarray*}
\Delta E({\bf p},{\bf k})&=&\epsilon({\bf p}+\hbar{\bf k})-\epsilon({\bf p})\\
&=&\hbar{\bf k}\cdot{\bf p}/m+\hbar^{2}k^{2}/2m\quad\text{(quantum)}\\
&=&\hbar{\bf k}\cdot{\bf p}/m \quad\text{(classical)}
\end{eqnarray*}
In the classical case, energy is always conserved only to lowest order in the momentum transfer ${\bf k}$.
In the quantum case, the leading term at very large $k$ is $\hbar^{2}k^{2}/2m$, independent of the momentum ${\bf p}$.
As a consequence, at $\omega=0$,  the energy-conserving delta functions in Eqs.(\ref{quantumImChi}) and (\ref{classicalImChi}) brings in the $k$-dependent factor $f(k/2)$ in the quantum case and the $k$-independent factor $F_{cl}(p=0)=1$ in the classical case.
Hence the additional term $f(k/2)$ in the Coulomb logarithm of real plasmas, often attributed in the litterature to electron degeneracy, arises from quantum diffraction.

\subsection{The $f$-sum rule} \label{f-sum-rule}

As a consequence of causality, a density-density response function $\chi(k,z)$ in a one-component system satisfies
\begin{eqnarray}
\chi(k,z)&=&\frac{1}{\pi}\int_{-\infty}^{\infty}{d\omega\frac{{\rm Im}\chi(k,\omega)}{\omega-z}}\/,
\end{eqnarray}
for any complex number $z$.
Because ${\rm Im}\chi(k,\omega)$ is odd with respect to $\omega$, the previous relation implies,
\begin{eqnarray}
\chi(k,z)&=&-\frac{1}{\pi}\sum_{n=0}^{\infty}{\frac{\omega_{2n+1}}{z^{2n+1}}} \label{highfrequencylimit}
\end{eqnarray}
where $\omega_{n}(k)$ is the frequency momenty of order $n$,
\begin{eqnarray}
\omega_{n}(k)=\int_{-\infty}^{\infty}{d\omega}{\,\omega^{n}\,{\rm Im}\chi_{i}(k,\omega)}
\end{eqnarray}
The first moment $\omega_{1}$, which carries information on the very short-time dynamics of the system, is {\it independent} of the
interparticle interactions (over very short time scales, the particles motion is purely kinetic) and is equal to the ideal gas value,
\begin{eqnarray}
\omega_{1}(k)=\int_{-\infty}^{\infty}{d\omega}{\,\omega\,{\rm Im}\chi^{(0)}(k,\omega)}=\frac{\pi n\/k^{2}}{m} \label{fsumrule}
\end{eqnarray}

\section{Derivation of Eq.(\ref{Trelaxation})} \label{appendix1}

In this appendix, we show how to obtain Eq.(\ref{Trelaxation}) from the more general result Eq.(\ref{dTdtgeneral2}).
To this end, we note that in most pratical applications $\alpha\equiv m_{e}T_{i}/m_{i}T_{e}<<1$ ($m_{e}/m_{i}<1./1815.<<1$ when $T_{e}=T_{i}$) and, as a consequence, the $\omega$-integral can be performed analytically by exploiting the $f$-sum rule.
Similar but not identical calculations were performed by Hazak et al. \cite{Hazak} to simplify the full FGR formula of $\nu_{ie}$, and by Boercker et al. \cite{Boercker} to simplify their extended Ziman formula for the electrical conductivity.
Througout the appendix we use the general properties of response functions recalled in appendix \ref{appendixChi}.

First, we rewrite the integrand of Eq.(\ref{dTdtgeneral2}) with the help of the following quantities,
\begin{eqnarray}
\chi_{ee}(k,\omega)&\equiv&\frac{\chi_{e}^{(0)}(k,\omega)}{1-u_{ee}(k)\chi_{e}^{(0)}(k,\omega)}=\frac{\chi_{e}^{(0)}(k,\omega)}{\epsilon_{e}(k,\omega)}\\
\chi_{ii}(k,\omega)&\equiv&\frac{\chi_{i}^{(0)}(k,\omega)[1-u_{ee}(k)\chi_{e}^{(0)}(k,\omega)]}{D(k,\omega)}\nonumber\\
&\equiv&\frac{\chi_{i}^{(0)}(k,\omega)}{\epsilon_{i}(k,\omega)} \label{Chi_ii} \/,
\end{eqnarray}
where $D$ and $u_{ee}$ are defined as in the main text.
$\chi_{ee}$ resembles to the response function of the interacting one-component electron gas (classical or quantum jellium model) while $\chi_{ii}$ is similar to the ion-ion density response function of the ions in an electron-ion plasma \cite{IchimaruVol1}.
Since $u_{ab}(k)=v_{ab}(k)(1-G_{ab}(k))$ is real, the imaginary parts of $\chi_{ee}$ and $\chi_{ii}$ are ${\rm Im}\chi_{ee}={\rm Im}\chi_{e}^{(0)}/|\epsilon_{e}|^{2}$ and,
\begin{eqnarray}
{\rm Im}\chi_{ii}=\frac{{\rm Im}\chi_{i}^{(0)}}{|\epsilon_{i}|^{2}}+u_{ei}u_{ie}{\rm Im}\chi_{ee}|\chi_{ii}|^{2}\/.
\end{eqnarray}
Therefore, the quantity ${\rm Im}\chi_{e}^{0}{\rm Im}\chi_{i}^{0}/|D|^{2}$ that appears in the integrand of Eq.(\ref{dTdtgeneral2}) writes
\begin{eqnarray}
\frac{{\rm Im}\chi_{e}^{0}{\rm Im}\chi_{i}^{0}}{|D|^{2}}&=&\frac{{\rm Im}\chi_{e}^{0}}{|\epsilon_{e}|^{2}}\frac{{\rm Im}\chi_{i}^{0}}{|\epsilon_{i}|^{2}} \label{intermediate}\\&=&{\rm Im}\chi_{ee}{\rm Im}\chi_{ii}-u_{ei}(k)u_{ie}(k)[{\rm Im}\chi_{ee}]^{2}|\chi_{ii}|^{2} \nonumber
\end{eqnarray}

Because $\alpha << 1$, ions are much slower than electrons and the ionic spectrum of fluctuations ${\rm Im}\chi_{ii}(k,\omega)$ resides on a very low-frequency range.
First, at small $k$, where collective effects predominate and $Y=\sqrt{m_{i}\beta_{i}/2}\omega/k<<1$, ${\rm Im}\chi_{ii}(k,\omega)$ rapidly vanishes as the frequency $\omega$ exceeds the ion plasma frequency $\omega_{pi}=\sqrt{4\pi n_{e}Ze^{2}/m_{i}}$, while ${\rm Im}\chi_{ee}(k,\omega)$ peaks at the electronic plasma frequency $\omega_{pe}=\sqrt{4\pi n_{e}e^{2}/m_{e}}=\sqrt{m_{i}/Zm_{e}}\omega_{pi}>>\omega_{pi}$.
Therefore the quantities in Eq.(\ref{intermediate}) vanish in this limit and do not contribute to the $\omega$ integral in Eq.(\ref{Trelaxation}).
Second, ${\rm Im}\chi_{ii}(k,\omega)$ drops to zero very rapidly as the phase velocity $\omega/k$ of the excitation exceeds the ion thermal velocity $\sqrt{k_{B}T_{i}/m_{i}}$, i.e. $Y=\sqrt{m_{i}\beta_{i}/2}\omega/k>1$, since in this limit ${\rm Im}\chi_{ii}(k,\omega)\approx{\rm Im}\chi_{i}^{(0)}(k,\omega)$ and
\begin{eqnarray}
{\rm Im}\chi_{i}^{(0)}(k,\omega)=-n_{i}\beta_{i}\sqrt{\pi}\/Ye^{-Y^{2}}\/.
\end{eqnarray}
Overall, the range of frequencies that contributes to the $\omega$ integral in Eq.(\ref{dTdtgeneral2}) is characterized by $Y\leq 1$ and therefore $\alpha Y<<1$.
In this limit, ${\rm Im}\chi_{ee}$ can be replaced by its low-frequency limit,
\begin{eqnarray}
{\rm Im}\chi_{ee}(k,\omega)&\approx& \omega \frac{\partial}{\partial \omega}{\rm Im}\chi_{ee}(k,0)\\
&=&\omega\frac{1}{|\epsilon_{e}(k,0)|^{2}}\frac{\partial}{\partial \omega}{\rm Im}\chi_{ee}^{(0)}(k,0)
\end{eqnarray}
since $\omega<<\omega_{pe}$ and ${\rm Im}\chi_{ee}(k,\omega)$ depends on $k$ and $\alpha Y$ according to (see Eq.(\ref{quantumImChi2}))
\begin{eqnarray*}
\lefteqn{{\rm Im}\chi_{e}^{(0)}(k,\omega)=}&&\\
&&=\frac{m_{e}^{2}k_{B}T_{e}}{2\pi\hbar^{4}k}\ln\frac{1+e^{-\left[\sqrt{\frac{1}{m_{e}k_{B}T_{e}}}\frac{\hbar k}{2}+\alpha Y\right]^{2}+\frac{\mu}{k_{B}T_{e}}}}{1+e^{-\left[-\sqrt{\frac{1}{m_{e}k_{B}T_{e}}}\frac{\hbar k}{2}+\alpha Y\right]^{2}+\frac{\mu}{k_{B}T_{e}}}}\\
&&\hspace{5cm}\text{(quantum)}\\
&&=-n_{e}\beta_{e}\sqrt{\pi}\/(\alpha Y)e^{-(\alpha Y)^{2}}\hspace{1.2cm}\text{(classical)}
\end{eqnarray*}
Accordingly, Eq.(\ref{intermediate}) becomes
\begin{eqnarray}
\lefteqn{\frac{{\rm Im}\chi_{e}^{(0)}(k,\omega){\rm Im}\chi_{i}^{(0)}(k,\omega)}{|D(k,\omega)|^{2}}}&&\nonumber\\
&\approx&\frac{\partial}{\partial \omega}{\rm Im}\chi_{ee}(k,0)\times\omega{\rm Im}\chi_{ii}(k,\omega) \label{intermediate2}\\
&&-u_{ie}(k)u_{ei}(k)\left(\frac{\partial}{\partial \omega}{\rm Im}\chi_{ee}(k,0)\right)^{2}\omega^{2}|\chi_{ii}(k,\omega)|^{2}\nonumber
\end{eqnarray}

Direct numerical calculations show that second term in the right-hand side of Eq.(\ref{intermediate2}) contributes negligibly to the Coulomb logarithm (systematically around $0.1\%$ of total Coulomb logarithm) and we neglect it from now on.

Keeping only the first term of Eq.(\ref{intermediate2}) in Eq.(\ref{dTdtgeneral2}), we obtain
\begin{eqnarray}
\frac{d T_{i}}{dt}&=&-\frac{2}{3k_{B}\pi n_{i}}\sqrt{\frac{m_{e}}{2k_{B}T_{e}}} \label{dTdtproof}\\
&\times&\int{\frac{d{\bf k}}{(2\pi)^{3}}v_{ie}(k)u_{ei}(k)\frac{\partial}{\partial \omega}{\rm Im}\chi_{ee}(k,\omega=0){\cal{S}}(k)}\nonumber
\end{eqnarray}
with 
\begin{eqnarray} 
{\cal{S}}(k)&=&\hbar\int{d\omega}{\omega^{2}\left[n\left(\frac{\hbar\omega}{k_{B}T_{i}}\right)-n\left(\frac{\hbar\omega}{k_{B}T_{e}}\right)\right]{\rm Im}\chi_{ii}(k,\omega)}\nonumber
\end{eqnarray}
For classical electrons, $n(x)=1/x$ and
\begin{eqnarray} 
{\cal{S}}(k)&=&k_{B}(T_{i}-T_{e})\omega_{1}(k) \label{Sk}
\end{eqnarray}
where
\begin{eqnarray}
\omega_{1}(k)&=&\int_{-\infty}^{\infty}{d\omega}{\omega{\rm Im}\chi_{ii}(k,\omega)} \label{omega_1}
\end{eqnarray}
For quantum electrons, $\hbar\omega/k_{B}T_{e,i}<<1$ over the low-frequency range outlined before, and therefore
\begin{eqnarray}
n\left(\frac{\hbar\omega}{k_{B}T_{i}}\right)-n\left(\frac{\hbar\omega}{k_{B}T_{e}}\right)\approx \frac{k_{B}T_{i}}{\hbar\omega}-\frac{k_{B}T_{e}}{\hbar\omega}
\end{eqnarray}
and ${\cal{S}}(k)$ is again given by Eq.(\ref{Sk}).

The moment $\omega_{1}(k)$ can be calculated exactly by noting that $\chi_{ii}$ satisfies the $f$-sum rule satisified by ordinary response functions (see appendix \ref{f-sum-rule}).
Indeed, at large frequency $\omega>>1$ and according to Eqs.(\ref{highfrequencylimit}) and (\ref{fsumrule}), 
\begin{eqnarray}
\chi_{i,e}^{(0)}(k,\omega)\approx \frac{\pi n_{e,i}k^{2}}{m_{i,e}\omega^{2}}\quad \omega>>1. \label{intermediatechi0highfrequency}
\end{eqnarray}
Substituting Eq.(\ref{intermediatechi0highfrequency}) in Eq.(\ref{Chi_ii}) implies,
\begin{eqnarray}
\chi_{ii}(k,\omega)\approx \frac{\pi n_{i}\/k^{2}}{m_{i}\omega^{2}}\quad\omega>>1 \label{w1_a}
\end{eqnarray}
If we assume that $\chi_{ii}$ is causal like its equilibrium counterpart (see the remark in appendix \ref{appendix2}), then according to Eq.(\ref{highfrequencylimit}),
\begin{eqnarray}
\chi_{ii}(k,\omega)\approx \omega_{1}(k)/\omega^{2}\quad \omega>>1 \label{w1_b}
\end{eqnarray}
and therefore, comparing Eqs.(\ref{w1_a}) and (\ref{w1_b}),
\begin{eqnarray}
\omega_{1}(k)=\frac{\pi n_{i}k^{2}}{m_{i}}\/, \label{f-sumrule}
\end{eqnarray}
Eq.(\ref{f-sumrule}) is valid for both quantum and classical electrons.
Direct numerical evaluation of Eq.(\ref{omega_1}) confirms the sum rule Eq.(\ref{f-sumrule}).

Finally, for both quantum and classical electrons, we find
\begin{eqnarray}
{\cal{S}}(k)=k_{B}(T_{i}-T_{e})\frac{\pi n_{i}k^{2}}{m_{i}}
\end{eqnarray}
and
\begin{eqnarray}
\frac{d T_{i}}{dt}&=&-\nu_{ie}\left(T_{i}-T_{e}\right)
\end{eqnarray}
where $\nu_{ie}$ is given by Eq.(\ref{model})

\begin{figure}[t]
\includegraphics[scale=0.50]{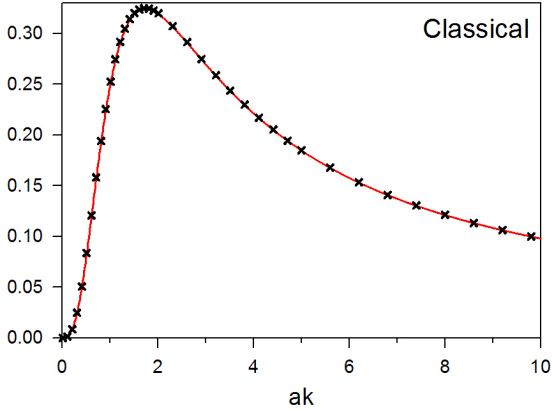}
\caption{\label{figure7}
Dimensionless $k$-integrand of Eq.(\ref{modellnL_a}) (full line) and Eq.(\ref{modellnL_b}) (crosses) for a classical, like-charged plasma obtained assuming $G_{\alpha\beta}=0$.}
\end{figure}

\begin{figure}[t]
\includegraphics[scale=0.50]{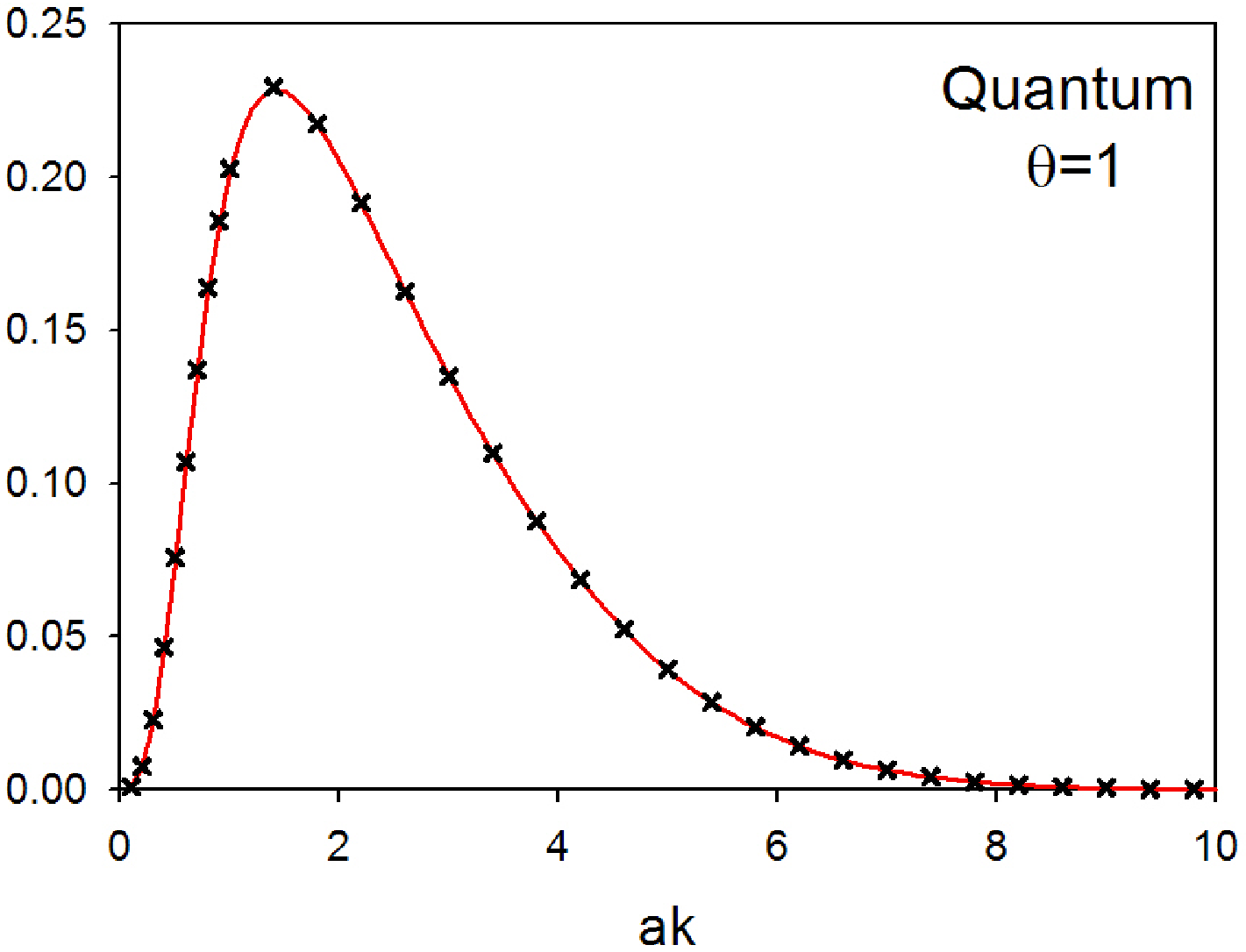}
\includegraphics[scale=0.50]{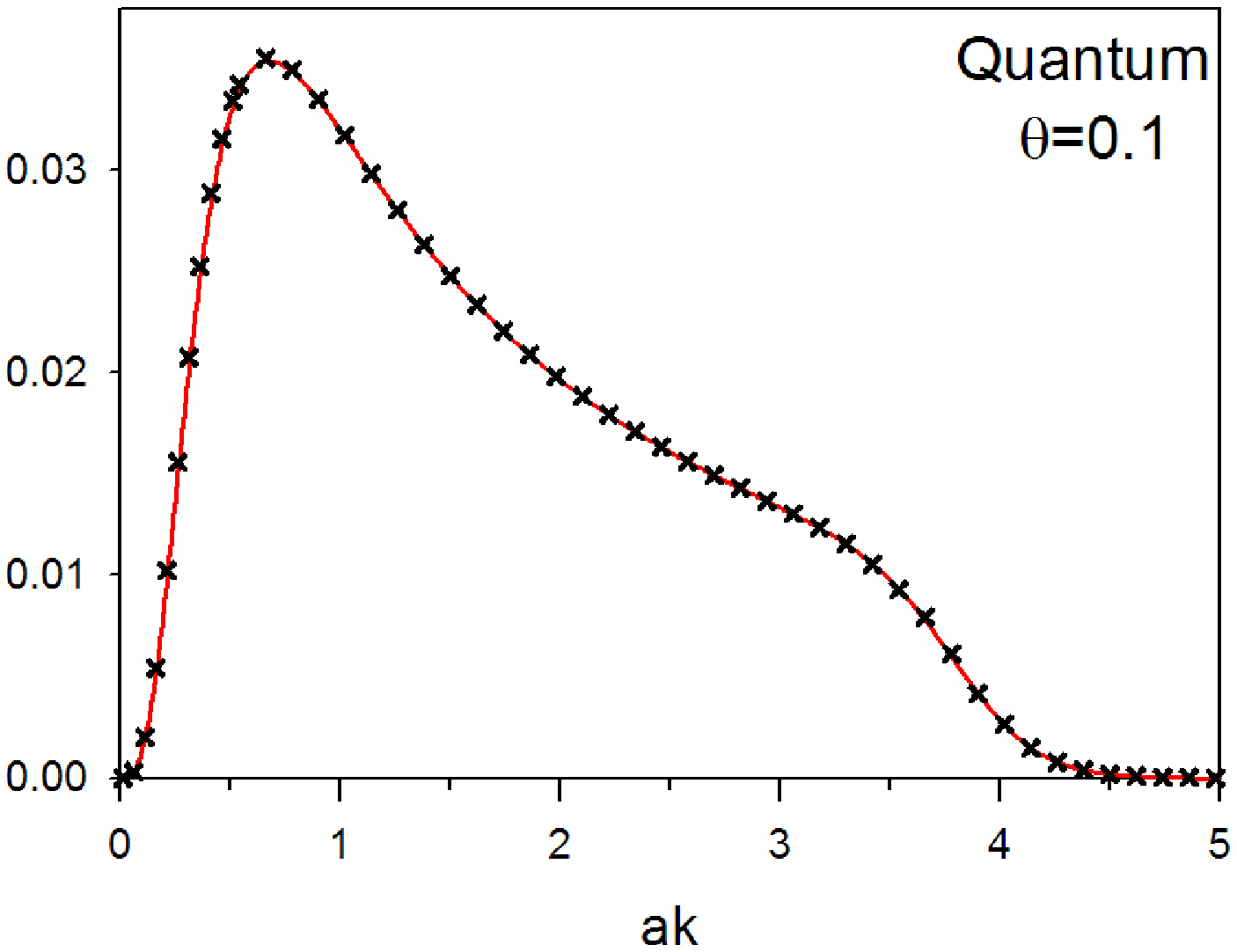}
\caption{\label{figure8}
Dimensionless $k$-integrand of Eq.(\ref{modellnL_a}) (full line) and Eq.(\ref{modellnL_b}) (crosses) for a hydrogen plasma at degeneracy $\Theta=1.$ (upper part) and $\Theta=0.1$ (lower part) and $\Gamma=1$.}
\end{figure}

The high accuracy of the small $\alpha=m_{e}T_{i}/m_{i}T_{e}$ expansion described here is illustrated in Figs.\ref{figure7} and \ref{figure8} for classical electrons and for quantum electrons with degeneracy $\Theta=1$ and $\Theta=0.1$, respectively.
Figs. \ref{figure7} and \ref{figure8} show a comparison between the $k$-integrands in the following two expressions for the Coulomb logarithm with $v_{ie}(k)=4\pi ZZ_{e}e^{2}/k^{2}$, namely
\begin{eqnarray}
\ln\Lambda&=&-\sqrt{\frac{2}{\pi m_{e}}}\frac{(k_{B}T_{e})^{3/2}}{n_{e}}\frac{m_{i}}{\pi n_{i}}\label{modellnL_a}\\
&&\times\int{dk \frac{1}{k^{2}}\left[1-G_{ie}(k)\right]\int{d\omega}{\frac{{\rm Im}\chi_{e}^{0}(k,\omega){\rm Im}\chi_{i}^{0}(k,\omega)}{|D(k,\omega)|^{2}}}}\nonumber\/,
\end{eqnarray}
obtained by calculating {\it numerically} the $\omega$ integral, and
\begin{eqnarray}
\ln\Lambda&=&-\sqrt{\frac{2}{\pi m_{e}}}\frac{(k_{B}T_{e})^{3/2}}{n_{e}} \label{modellnL_b}\\
&&\times\int_{0}^{\infty}{\!\!\!\!dk \left|\frac{1\!-\!G_{ie}(k)}{\epsilon_{e}(k,0)}\right|^{2}\frac{\partial {\rm Im}\chi_{e}^{0}(k,\omega)}{\partial\omega}\Big|_{\omega=0}}\/, \nonumber
\end{eqnarray}
obtained as before by {\it exploiting the $f$-sum rule}.
As expected, both calculations agree very well.

\section{$\lim_{k\to\infty}{1-G_{ie}(k)}$} \label{appendixlimGie}

In this appendix, we prove the results (\ref{limGieclassical}) and (\ref{limGiequantum}) used in the main text.
To this end, we combine the OZ relation (\ref{Gie}) valid for both classical and quantum electron with:
\begin{itemize}
\item the asymtotic limit of $S_{ie}(k)$:
\begin{eqnarray*}
S_{ie}(k)&=&\sqrt{n_{e}n_{i}}\int{d{\bf r}\left[g_{ie}(r)-1\right]e^{-i{\bf k}\cdot{\bf r}}}\\
&=&\int_{0}^{\infty}{dr f(r) e^{-ikr}}\\
&=&\frac{4\pi\sqrt{n_{e}n_{i}}}{k}\left[-\frac{2g_{ie}^{\prime}(0)}{k^{3}}+\frac{4g_{ie}^{\prime\prime\prime}(0)}{k}+\dots\right]\quad k\sim\infty
\end{eqnarray*}
and therefore, using the cusp condition at the origin $g_{ie}^{\prime}(0)=-2g_{ie}(0)/a_{B}$ \cite{NagaoBonevAshcroft} for real plasmas and $g_{ie}^{\prime}(0)=0$ for liked-charged plasmas,
\begin{eqnarray*}
S_{ie}(k)\sim \left\{
\begin{array}{l}
\frac{16\pi n_{e}}{\sqrt{Z}a_{B}}\frac{1}{k^{4}}g_{ie}(r=0)\quad\text{(quantum)}\nonumber\\
\\
\frac{16\pi}{k^{6}}g^{(3)}(r=0)\quad\text{(classical)}\nonumber\\
\end{array}\right.
\hspace{1cm}k\sim\infty
\end{eqnarray*}
\item the asymptotic behavior of $\chi_{e}^{(0)}(k,0)$: using the results of appendix \ref{appendixChi},
\begin{eqnarray*}
\frac{\chi_{e}^{(0)}(k,0)}{\chi_{e}^{(0)}(0,0)}\sim\left\{
\begin{array}{l}
\frac{1}{k^{2}}\quad\text{(quantum)}\nonumber\\
\\
1\quad\text{(classical)}\nonumber\\
\end{array}\right.
\hspace{1cm}k\sim\infty
\end{eqnarray*}
\item the asymptotic limit of $D(k,0)$:
\begin{eqnarray*}
D(k,0)\sim 1\hspace{1cm}k\sim \infty
\end{eqnarray*}
\end{itemize}
Combining these results in Eq.(\ref{Gie}), we obtain
\begin{eqnarray}
1-G_{ie}(k)\left\{
\begin{array}{l}
\propto g_{ie}(r=0)>0 \quad\text{(quantum)}\nonumber\\
\\
=0\quad\text{(classical)}\nonumber\\
\end{array}\right.
\hspace{1cm}k\sim\infty
\end{eqnarray}

\section{Comparison with other models} \label{comparison}

In this appendix, we provide a comparison between our model and three other models, namely the ``Fermi-Golden Rule'' (FGR) model described by Hazak et al. in \cite{Hazak} and also by \cite{DharmaWardanaPerrot}, the Lenard-Balescu (LB) model discussed e.g. in \cite{Gericke} and the Dharma-wardana and Perrot (DWP) model developed in \cite{DharmaWardanaPerrot}.

To this end, the four models are conveniently expressed into similar-looking expressions for the evolution of the energy density ${\cal{E}}_{i}=3n_{i}k_{B}T_{i}/2$ as
\begin{widetext}
\begin{eqnarray}
\!\!\!\frac{d{\cal{E}}_{i}}{dt}&\!\!=\!\!&-2\hbar\int{\!\!\frac{d{\bf k}}{(2\pi)^{3}}\int{\!\!\frac{d\omega}{2\pi}\big|v_{ie}(k)\big|^{2}}}\omega\Delta N(k,\omega){\rm Im}\chi_{e}^{ocp}(k,\omega){\rm Im}\chi_{i}^{ocp}(k,\omega)\hspace{1.3cm}\text{Fermi Golden Rule \cite{Hazak,DharmaWardanaPerrot}} \label{FGR}\\
\nonumber\\
&\!\!=\!\!&-2\hbar\int{\!\!\frac{d{\bf k}}{(2\pi)^{3}}\int{\!\!\frac{d\omega}{2\pi}\big|v_{ie}(k)\big|^{2}}}\omega\Delta N(k,\omega)\frac{{\rm Im}\chi_{e}^{lb}(k,\omega){\rm Im}\chi_{i}^{lb}(k,\omega)}{|1-|v_{ei}(k)|^{2}\chi_{e}^{lb}(k,\omega)\chi_{i}^{lb}(k,\omega)|^{2}}\hspace{0.3cm}\text{Lenard-Balescu \cite{Gericke}} \label{LenardBalescu}\\
\nonumber\\
&\!\!=\!\!&-2\hbar\int{\!\!\frac{d{\bf k}}{(2\pi)^{3}}\int{\!\!\frac{d\omega}{2\pi}\big|v_{ie}(k)\big|^{2}}}\omega\Delta N(k,\omega)\frac{{\rm Im}\chi_{e}^{dp}(k,\omega){\rm Im}\chi_{i}^{dp}(k,\omega)}{|1-|v_{ei}(k)|^{2}\chi_{e}^{dp}(k,\omega)\chi_{i}^{dp}(k,\omega)|^{2}}\hspace{0.2cm}\text{Dharma-wardana-Perrot \cite{DharmaWardanaPerrot}} \label{DWP}\\
\nonumber\\
&\!\!=\!\!&-2\hbar\int{\!\!\frac{d{\bf k}}{(2\pi)^{3}}\int{\!\!\frac{d\omega}{2\pi}\big|v_{ie}(k)\big|^{2}}}\omega\Delta N(k,\omega)\frac{\left[1-G_{ie}(k)\right]{\rm Im}\chi_{e}(k,\omega){\rm Im}\chi_{i}(k,\omega)}{|1-|v_{ei}(k)|^{2}(1-G_{ei}(k))(1-G_{ie}(k))\chi_{e}(k,\omega)\chi_{i}(k,\omega)|^{2}}\nonumber\\
&&\hspace{11.cm}\text{Our model, Eq.(\ref{dTdtgeneral2})} \label{ourmodel}
\end{eqnarray}
\end{widetext}
with $\Delta N(k,\omega)\equiv n(\hbar\omega/k_{B}T_{i})-n(\hbar\omega/k_{B}T_{e})$.

Equation (\ref{ourmodel}) is nothing but our model Eq.(\ref{dTdtgeneral2}) written in terms of the quantities
\begin{eqnarray}
\chi_{e}(k,\omega)&\equiv&\frac{\chi_{e}^{(0)}(k,\omega)}{1-v_{ee}(k)(1-G_{ee}(k))\chi_{e}^{(0)}(k,\omega)}\\
\chi_{i}(k,\omega)&\equiv&\frac{\chi_{i}^{(0)}(k,\omega)}{1-v_{ii}(k)(1-G_{ii}(k))\chi_{i}^{(0)}(k,\omega)}\/.
\end{eqnarray}
Indeed, using ${\rm Im}\chi_{\alpha}={\rm Im}\chi_{\alpha}^{(0)}/|1-v_{\alpha}(k)(1-G_{\alpha\alpha})\chi_{\alpha}^{(0)}|^{2}$ in  Eq.(\ref{dTdtgeneral2}) leads to Eq.(\ref{ourmodel}).
Recall that here $G_{ee}$ and $G_{ii}$ are the static LFC's of a two-temperature, {\it two-component} electron-ion plasma introduced in the main text.

The FGR model (\ref{FGR}) was obtained by calculating the energy transfers in the first Born approximation in the e-i interaction.
As a result, in (\ref{FGR}), $\chi_{e,i}^{ocp}$ are the response functions of the interacting one-component electron and ion plasmas,
\begin{eqnarray}
\!\!\chi_{e}^{ocp}(k,\omega)&\!\!\equiv\!\!&\frac{\chi_{e}^{(0)}(k,\omega)}{1-v_{ee}(k)(1-G_{ee}^{ocp}(k,\omega))\chi_{e}^{(0)}(k,\omega)}\\
\!\!\chi_{i}^{ocp}(k,\omega)&\!\!\equiv\!\!&\frac{\chi_{i}^{(0)}(k,\omega)}{1-v_{ii}(k)(1-G_{ii}^{ocp}(k,\omega))\chi_{i}^{(0)}(k,\omega)}
\end{eqnarray}
where $G_{ee}^{ocp}$ and $G_{ii}^{ocp}$ are the LFC's for the {\it one-component} electron and ion gas at temperature $T_{e}$ and $T_{i}$, respectively.

The LB model (\ref{LenardBalescu}) is obtained from momentum integration of the LB kinetic equations \cite{Gericke}.
The response functions in Eq.(\ref{LenardBalescu}) are
\begin{eqnarray}
\chi_{e}^{lb}(k,\omega)&\equiv&\frac{\chi_{e}^{(0)}(k,\omega)}{1-v_{ee}(k)\chi_{e}^{(0)}(k,\omega)}\\
\chi_{i}^{lb}(k,\omega)&\equiv&\frac{\chi_{i}^{(0)}(k,\omega)}{1-v_{ii}(k)\chi_{i}^{(0)}(k,\omega)}
\end{eqnarray}

Comparing our model with the LB and FGR models, we remark that
\begin{itemize}
\item The physical interpretation of the three models goes as follows.
The FGR formula describes the energy exchanges between {\it two} weakly interacting, one-component electron and ion plasmas.
The two subsystems are independent from each other.
For instance, electrons do not affect the inter-ionic interactions; in reality, however, electrons screen the i-i interactions and the ionic plasmon excitation becomes an ion-acoustic mode.
Conversely, both the LB model and our model treat the entire plasma, i.e. electrons plus ions, as a {\it single} system, in which the collisions are due to the interactions via an effective, screening potential.
In both models the effective potential is not an {\it ad hoc} construct but is derived from a model like (\ref{ourmodel}) for the density fluctuations  $\delta n_{e,i}$ in the plasma.
In the LB model, the plasma is assumed to be weakly coupled and the density fluctuations $\delta n_{e,i}$ are accordingly described at the level of the random-phase approximation (RPA) \cite{IchimaruVol1}, which amounts to neglect the LFC's in Eq.(\ref{DensityFluctuations}) and therefore in Eq.(\ref{ourmodel}).
It is therefore not surprising that our model reduces to the LB result (\ref{LenardBalescu}) by setting the LFC's $G_{\alpha\beta}$ to zero in Eq.(\ref{ourmodel}).
In our model, the effect of particle correlations neglected in the RPA approximation are modeled using LFC's in Eq.(\ref{DensityFluctuations}).
Correlations modify the LB model in into two major ways.
First, they modify the dispersion relation of the collective modes in the plasma, i.e. the poles in Eqs.(\ref{LenardBalescu}) and (\ref{ourmodel}).
Howvever, as discussed in the main text and below, this effect barely affects the relaxation rate when $m_{e}T_{i}/m_{i}T_{e}<<1$ since in this limit the e-i energy exchanges are globally insensitive to the details of the fluctuation spectrum.
Second, they account for the short-range e-i correlations ($g_{ei}(r=0)>1$ and therefore $1-G_{ie}(k=\infty)\neq 0$ in real plasma, $g_{ie}(0)=0$  and therefore $1-G_{ie}(k=\infty)=0$ in like-charged plasmas) and bring in the contribution of close encounters (as discussed in the main text, in real plasma, both short-range correlations and quantum diffraction determine the effect of close encounters.)
\item From appendix \ref{appendix1}, it is clear that in the limit $m_{e}T_{i}/m_{i}T_{e}<<1$ both the LB and FGR models
\begin{eqnarray}
\lefteqn{\ln\Lambda=}&&\label{FGRsmallalpha}\\
&&\int_{0}^{\infty}{\frac{dk}{k}\frac{1}{|1-v_{ee}(k)[1-G(k,0)]\chi^{0}_{e}(k,0)|^{2}}\/f(k/2)}\nonumber
\end{eqnarray}
with $G=0$ for LB and $G=G_{ee}^{ocp}(k,0)$ for FGR.
Eq.(\ref{FGRsmallalpha}) was previously derived and validated in \cite{Hazak} for the FGR model and in \cite{DaligaultMozyrsky} for the LB model.
As discussed in the main text, our model reduces to Eq.(\ref{FGRsmallalpha}) with $G=G_{ee}$ when the e-i correlation effects are neglected, i.e. $G_{ie}=0$ in Eqs. (\ref{classicalmodel}) or (\ref{quantummodel}).
The three models agree when all the LFC's are neglected.
\item As discussed in the main text, Eq.(\ref{FGRsmallalpha}) diverges when electrons are treated classically (i.e. when $f(k/2)\equiv1$ in Eq.(\ref{FGRsmallalpha})) because of the inadequate treatment of close e-i encounters.
Short-range correlation effects described by the $1-G_{ie}$ term in Eq.(\ref{classicalmodel}) are crucial in purely classical plasmas at any density and temperature to provide the large momentum cutoff.
The LB and FGR models do not describe those correlations properly and accordingly diverge at large $k$ for classical electrons.
\item Eq.(\ref{FGRsmallalpha}) converges when electrons are treated quantum mechnically, because $f(k/2)$ vanishes exponentially at large $k$ and cuts off the integral for $k$ of the order of $2k_{F}\sqrt{1+\Theta}$.
With $G\equiv 0$, all three models lead to the Coulomb logarithm,
\begin{eqnarray}
\ln\Lambda&=&\int_{0}^{\infty}{dk\frac{k^{3}}{(k^{2}+1)^{2}}e^{-\frac{\lambda_{th}^{2}}{8\lambda_{De}^{2}}k^{2}}} \nonumber\\
&\approx&\ln(0.742\sqrt{r_{s}}/\Gamma)\quad \Gamma<<1 \label{nuLB}
\end{eqnarray}
Note that BPS \cite{BPS} also recover the same result in the so-called quantum limit $r_{L}<\lambda_{th}$; see also \cite{LandauLifshitz}.
\end{itemize} 

The DWP theory (\ref{DWP}) was proposed as the non-perturbative extension of the FGR theory.
DWP attribute the energy relaxation to the dynamically coupled electron and ion modes in the plasma resulting from the strong e-i interactions.
These modes are accounted for by the dynamical response functions $\chi_{e,i}^{dp}(k,\omega)$ discussed below.
At melting temperature, the DWP relaxation rates predicted for metals are an order of magnitude smaller than the FGR rates \cite{DharmaWardanaPerrot2}.
For hot dense hydrogen, according to the results recently reported in \cite{MurilloDharmaWardana} for densities $r_{s}=0.5$ and $r_{s}=1$ and temperatures $50<T<4000$ $\rm eV$ (see their Fig.2), the DWP relaxation rates are systematically lower than the FGR rates, both rates are close (but not equal !) at temperatures $T\ge 1$ $\rm keV$ and the DWP model deviates non-negliglibly and increasingly at lower temperatures.
These findings differ from the conclusions of our model described in the main text.

The description and the notations in \cite{DharmaWardanaPerrot,DharmaWardanaPerrot2,MurilloDharmaWardana} are rather confusing concerning the definition of the quantities $\chi_{e,i}^{dp}(k,\omega)$ to be used in Eq.(\ref{DWP}).

According to the recent paper \cite{MurilloDharmaWardana}, the DWP model is simply obtained by including the denominator $1-v_{ie}^{2}\chi_{e}^{ocp}\chi_{i}^{ocp}$ in the integrand (\ref{FGR}) of the FGP formula (see Eq.(9) and related text in \cite{MurilloDharmaWardana}), i.e. $\chi_{e,i}^{dp}=\chi_{e,i}^{ocp}$ in Eq.(\ref{DWP}).
This seems unlikely since in that case the DWP model (\ref{DWP}) would be nothing more than the LB model (\ref{LenardBalescu}) corrected by the LFC $G_{ee,ii}^{ocp}$ (in practice, \cite{DharmaWardanaPerrot,DharmaWardanaPerrot2,MurilloDharmaWardana} use static LFC).
Then, just like the LB and FGR models, DWP would also be well approximated by Eq.(\ref{FGRsmallalpha}) and would not significantly differ from FGR.

According to a previous paper \cite{DharmaWardanaPerrot2}, the $\chi_{e,i}^{dp}$ in Eq.(\ref{DWP}) are the electron and ion response functions of a two-component plasma, namely
\begin{eqnarray*}
\chi_{\alpha}^{dp}&=&\chi_{\alpha}^{(0)}\left[1-v_{\alpha\alpha}(1-G_{\alpha\alpha})\chi_{\alpha}^{(0)}\right]/D\quad\alpha=e,i\\
D&=&\left[1-v_{ee}(1-G_{ee})\chi_{e}^{(0)}\right]\left[1-v_{ii}(1-G_{ii})\chi_{i}^{(0)}\right]\nonumber\\
&&\quad-v_{ei}v_{ie}(1-G_{ei})(1-G_{ie})\chi_{e}^{(0)}\chi_{i}^{(0)}
\end{eqnarray*}
With these definitions, it is hard to see how to perform analytically the $\omega$-integral in Eq.(\ref{DWP}) as is possible with all the other models using the $f$-sum rule.
Indeed, since $\chi_{e}^{dp}$ is sensitive to the low frequency (ion-acoustic) mode, the quantity ${\rm Im}\chi_{e}^{dp}{\rm Im}\chi_{i}^{dp}/|1-|v_{ei}|^{2}\chi_{e}^{dp}\chi_{i}^{dp}|^{2}$ cannot easily be written as the product of two quantities, one that resides on a very low-frequency range over which the other barely varies.
Accordingly, the DWP theory is sensitive to the details of the ion fluctuation spectrum and find results that differ from FGR in the way recalled above.

Comparing our model with DWP, we remark that:
\begin{itemize}
\item as discussed earlier, both our model and the LB model describe self-consistently the electron-ion interactions in a plasma and therefore the coupled modes at the root of DWP theory.
The LB model describes those modes at the level of the RPA approximation while our model brings in correlation effects.
Again, because in practice electrons are much faster than ions, $\alpha=m_{e}T_{i}/m_{i}T_{e}<<1$, the temperature relaxation rates are insensitive to the mode spectrum and both the LB model and our model reduce to Eq.(\ref{FGRsmallalpha}).
We suspect that the prescription of DWP, supposedly based on some of the Feymann diagrams discussed in \cite{DharmaWardanaPerrot}, overestimates (multiple counts) the effect of screening.
\item unlike our model, the DWP expression (\ref{DWP}) does not include a term like $1-G_{ie}(k)$ in the numerator.
As a consequence, when applied to a classical like-charged system, the DWP model {\it diverges} logarithmically at large-k (the integrand scale like $1/k$ at large $k$ as the LB and FGR formulas).
The DWP is therefore unable to explain the classical MD data discussed in section \ref{section3}, while the concept of coupled-modes at the basis of DWP does not only pertain to real plasmas but also applies to classical like-charged systems.
\item for dense hydrogen, our model predict rates at most $50\%$ larger than the FGR, in contrast with the findings of \cite{MurilloDharmaWardana} mentionned above.
\end{itemize}


\begin{references}
\bibitem{Noteonpseudopotential} The e-i interaction potential can be modeled with the bare Coulomb interaction $v_{ie}(k)=4\pi ZZ_{e}e^{2}/k^{2}$ or with a pseudopotential when a nucleus and its tightly bound electrons behave as a rigid entity.
\bibitem{NoteDetailedBalance} When electrons are treated quantum mechanically, the quantum expression for $S_{ii}^{0}$ is used in order to satisfy detailed balance and ultimately energy conservation.
\bibitem{NoteOZ} This equation is easily obtained by rewriting of the OZ relation for a two-component system in terms of the structure factors and the local field corrections (recall that at equlibrium the static LFC, say $G_{ie}$, and the direct correlation function $c_{ie}$ are simply related such as $c_{ie}(k)=-v_{ie}(1-G_{ie}(k))/k_{B}T$.) Alternatively, Eq.(\ref{Gie}) can be obtained from the flucutation-dissipation theorem relating the struture factors to the response functions.
\bibitem{noteHNC} Here we describe calculations obtained using the HNC equations with $T_{e}=T_{i}$. Results with $T_{e}\neq T_{i}$ were not significantly different.
\bibitem{NotekTF} We use $\epsilon_{e}(k)\approx 1-4\pi\/e^{2}{\rm Re}\chi_{e}^{(0)}(k,0)/k^{2}\approx 1+k_{TF}^{2}/k^{2}$ where $k_{TF}=(-4\pi e^{2}{\rm Re}\chi_{e}^{(0)}(0,0))^{1/2}=4k_{F}\sqrt{\Theta}I_{1/2}^{\prime}(\eta)/\pi\/a_{B}$ is the finite temperature Thomas-Fermi screening length (here $I_{1/2}$ is the usual Fermi function so that $n_{e}=I_{1/2}(\eta)/2\pi^{2}\hbar^{3}(2m_{e}k_{B}T_{e})^{3/2}$) \cite{IchimaruVol1,Kremp}.
\end{references}
\end{document}